\documentclass[usegraphicx,useAMS]{mn2e}
\begin{document}

\title[The HIDEEP Survey]{HIDEEP - An extragalactic blind survey for very low 
column-density neutral hydrogen}

\author[R. F. Minchin et al.]{R.~F.~Minchin,$^1$ M.~J.~Disney,$^1$
P.~J.~Boyce,$^2$ W.~J.~G.~de~Blok,$^1$, Q.~A.~Parker,$^3$ 
\newauthor G.~D.~Banks,$^1$\thanks{Now at BAE Systems}
K.~C.~Freeman,$^4$ D.~A.~Garcia,$^1$ M.~Grossi,$^1$ R.~F.~Haynes,$^5$ 
\newauthor P.~M.~Knezek,$^6$ R.~H.~Lang,$^1$ D.~F.~Malin,$^7$ 
R.~M.~Price,$^{8}$ I.~M.~Stewart,$^{9}$ 
\newauthor A.~E.~Wright$^{10}$\\
$^1$ School of Physics and Astronomy, Cardiff University, 5 The Parade, 
Cardiff, CF24 3YB\\
$^2$ Astrophysics Group, Department of Physics, University of Bristol, Tyndall
Avenue, Bristol, BS8 1TL \\
$^3$ Department of Physics, Macquarie University, Sydney, NSW 2109, Australia\\
$^4$ Research School of Astronomy \& Astrophysics, Mount Stromlo Observatory, 
Cotter Road, Weston ACT 2611, Australia\\
$^5$ School of Mathematics \& Physics, University of Tasmania, Hobart, 
Tasmania 7001, Australia\\
$^6$ WIYN Consortium Inc., 950 North Cherry Avenue, Tucson, AZ 85719, United 
States\\
$^7$ Anglo-Australian Observatory, P.O. Box 296, Epping, NSW 1710, Australia\\
$^{8}$ Department of Physics and Astronomy, University of New Mexico, 
800 Yale Boulevard NE, Albuquerque, NM, United States\\
$^{9}$ Department of Physics and Astronomy, University of Leicester, 
University Road, Leicester LE1 7RH\\
$^{10}$ Australia Telescope National Facility, PO Box 76, Epping, NSW 1710, 
Australia}
\maketitle

\begin{abstract}
We have carried out an extremely long integration-time (9000s\,beam$^{-1}$)
21-cm blind survey of 60 square degrees in Centaurus using the 
Parkes multibeam system.  We find that the noise continues to fall as
$\sqrt{t_{obs}}$ throughout, enabling us to reach an H{\sc i} column-density 
limit of $4.2 \times 10^{18}$~cm$^{-2}$ for galaxies with a velocity width 
of 200~km\,s$^{-1}$ in the central 32 square degree region,
making this the deepest survey to date in terms of column density sensitivity.
The H{\sc i} data are complemented by very deep optical observations from 
digital stacking of multi-exposure UK Schmidt Telescope {\it R}-band films, 
which reach an isophotal level of 26.5~$R$\,mag\,arcsec$^{-2}$ ($\simeq 
27.5$~$B$\,mag\,arcsec$^{-2}$). 173 H{\sc i} sources have been found, 96 of
which have been uniquely identified with optical counterparts in the overlap
area.  There is not a single source without an optical counterpart.  Although
we have not measured the column-densities directly, we have inferred them
from the optical sizes of their counterparts.  All appear to have a 
column-density of $N_{HI} = 10^{20.65\pm 0.38}$.  This is at least an order
of magnitude above our sensitivity limit, with  a scatter only 
marginally larger than the errors on $N_{HI}$.  This needs explaining.  If 
confirmed it 
means that H{\sc i} surveys will only find low surface brightness (LSB) 
galaxies with high $M_{HI}/L_B$.  Gas-rich LSB galaxies with lower H{\sc i} 
mass to light ratios do not exist.  The paucity of low column-density galaxies
also implies that no significant population will be missed by the all-sky
H{\sc i} surveys being carried out at Parkes and Jodrell Bank.

\end{abstract}
\section{introduction}
\label{intro-sec}

Most of our knowledge of galaxy populations derives from samples collected
 in optical surveys, where there are strong selection effects.
For instance most of the galaxies we know about are barely brighter than 
the terrestrial sky, while galaxies which have lower surface-brightnesses
may be severely under-represented. Attempts to overcome optical 
selection effects through the use of better detectors, larger
telescopes, longer exposures and sophisticated image processing have met
with partial success.  This remains a very difficult process, however, and the 
subsequent corrections which have to be made to small number statistics
are both large and controversial (e.g.\ Impey \& Bothun 1997, Disney 1999).
                           
Low surface brightness (LSB) galaxies could
be significant in a number of contexts.   They might add to or even dominate
the luminosity-density and/or mass-density of galaxies as a whole (e.g. 
Fukugita, Hogan, \& Peebles 1998).  Additionally, they could be responsible 
for much of the metal line absorption seen in QSO spectra (Churchill \& Le
 Brun 1998), while their presence in large numbers
could be important to theories of galaxy formation and evolution.
                       
Contemporary views as to the global significance of LSB galaxies are wide 
ranging, partly as a result of conflicting and usually indirect 
evidence, but mainly as a consequence of conflicting interpretation
(Impey \& Bothun 1997; Davies, Impey, \& Phillipps 1999).
                             
Blind searches in the 21-cm neutral hydrogen line have long been considered
an alternative to optical surveys for finding LSB disc galaxies (Disney 1976).
The narrowness of the line means that cosmic expansion will discriminate 
between extra-galactic and local hydrogen, so reducing the local background to
the instrumental level beyond a redshift of a few hundred km\,s$^{-1}$. 
Unfortunately, blind 21-cm surveys have until recently been severely limited 
by technical considerations, in particular by the tiny areas and small 
velocity ranges that could be covered to any depth in a practical time. 

\begin{table}
\caption{Equivalent central surface-brightnesses (in {\it B}-band)
for different column densities and different H{\sc i} mass-to-light ratios
(calculated using the derivation in Appendix \ref{disney-banks-deriv}).
It can be seen that if $M_{HI}/L_B$ remains constant at the value for
optically-selected galaxies of $\simeq 0.3$, it is necessary to
go to very low column-densities to find LSB galaxies.  However, if 
$M_{HI}/L_B$ rises as the surface-brightness falls, it is possible to 
reach low surface-brightnesses without requiring low column-densities.}
\label{nhi-sb_tab}
\begin{tabular}{lccccc}
&\multicolumn{5}{c}{$M_{HI}/L_{\rm B} \left(M_\odot / 
L_\odot\right)$}\\
$N_{HI}$ (cm$^{-2}$)&0.1&0.3&1&3&10\\
\hline
$10^{21}$&18.7&19.9&21.2&22.4&23.7\\
$3\times 10^{20}$&19.9&21.2&22.4&23.7&24.9\\
$10^{20}$&21.2&22.4&23.7&24.9&26.2\\
$3\times 10^{19}$&22.4&23.7&24.9&26.2&27.4\\
$10^{19}$&23.7&24.9&26.2&27.4&28.7\\
$3\times 10^{18}$&24.9&26.2&27.4&28.7&29.9\\
$10^{18}$&26.2&27.4&28.7&29.9&31.2\\
\end{tabular}
\end{table}

Recent technical 
advances, including high sensitivity multibeam receivers and powerful 
correlators, have allowed much more ambitious blind surveys to be carried out.
All-sky surveys of the southern hemisphere from Parkes (the H{\sc i} Parkes 
All Sky Survey, HIPASS) and of the northern hemisphere from Jodrell Bank
(the H{\sc i} Jodrell All Sky Survey, HIJASS) are currently being completed. 
These have 
integration times of 450s and 350s per beam respectively (Staveley-Smith et al.
1996; Lang et al. 2003), giving very similar sensitivities to sources smaller 
than the beam.  To supplement these shallow surveys we have carried out the 
much deeper HIDEEP survey (9000s\,beam$^{-1}$) of a small area of sky 
($4^\circ\times 8^\circ$) out to a velocity of 12,700 km\,s$^{-1}$.  

The main motive for HIDEEP was to reach previously inaccessible 
surface-brightness levels.  In general one might expect 
surface-brightness to 
be correlated with H{\sc i} column-density.  Indeed, if one assumes that 
gas and starlight are distributed over proportionate areas it is easy to show 
that (Appendix \ref{disney-banks-deriv}):
\begin{equation}
N_{HI} \simeq 10^{20.1}\left(\frac{M_{HI}}{L_B}\right)
10^{\left(0.4\left(27-\mu_{mean}(B)\right)\right)} 
\label{disney-banks}
\end{equation}

\noindent a relationship tabulated in Table \ref{nhi-sb_tab}.  To reach 
LSB galaxies ($\mu_{0} \geq 24$ {\it B}\,mag\,arcsec$^{-2}$) which have 
`normal' amounts of H{\sc i} (e.g. $M_{HI}/L_B = 0.3$) requires an H{\sc i} 
survey reaching down to $N_{HI} \leq 3\times 10^{19}$~cm$^{-2}$, while to 
reach really low surface-brightness objects with $\mu_0 \geq
26.5$ {\it B}\,mag\,arcsec$^{-2}$ requires an H{\sc i} survey ten times more 
sensitive.  Unfortunately, such low column-densities can be reached
 only by 
very long integrations, irrespective of dish size (as diffraction-limited 
beams decrease in angular area exactly in inverse proportion to dish area).  
In single-dish surveys, the detection of sources smaller than the beam 
will depend only on the mass of H{\sc i} ($M_{HI}$) per velocity channel
in the beam, and their H{\sc i} column-densities ($N_{HI}$, measured in 
cm$^{-2}$) will be irrelevant.
As we shall demonstrate (Section \ref{hideepsurvey} below) our
survey is peak-flux limited so that, for detection:

\begin{equation}
S_{peak} > n\sigma
\end{equation}

\noindent where $S_{peak}$ is the peak flux and $\sigma$ is the noise
per channel per beam.  As $S_{peak} \propto (M_{HI}/d^2\Delta V)$
and $\sigma \propto 1/(D^2t^{1/2})$, where $D$ is the diameter
of the dish, $d$ is the distance to the source, and $t_{int}$ is the
integration time, this can be re-written as:

\begin{equation}
\frac{1}{d^2}\left(\frac{M_{HI}}{\Delta V}\right) D^2 t^{1/2} > k_1
\label{diseqn0}
\end{equation}

\noindent where $k_1$ is a constant.  Thus the maximum distance for source 
detection will be:

\begin{equation}
d_{max} \propto \left[\left(\frac{M_{HI}}{\Delta V}\right)D^2
t^{1/2}\right]^{1/2}
\label{diseqn1}
\end{equation}

\noindent and the maximum volume in which such sources ($M_{HI}$, $\Delta V$)
will be detected is

\begin{equation}
V_{max} = \frac{\Omega_t}{3}N_b\times d_{max}^3 = 
\frac{\Omega_b}{3}\left(\frac{T}{t}\right)d_{max}^3
\end{equation}

\noindent (where $\Omega_t$ is beam size in sterads, $N_b$ is the total
number of beams in the survey and T is the total duration of the survey).

Thus the volume surveyed (and hence number of detections) per unit time is

\begin{equation}
\frac{V_{max}}{T} \propto \frac{1}{t}t^{3/4} \propto t^{-1/4}
\end{equation}

\noindent i.e. short integration times per beam are favoured in order to find
the most sources.

However, short integration times imply that sources
will only be detected nearby, and if they are too close they will over-fill
the beam, reducing the amount of H{\sc i} within it.  In these circumstances, 
Equation \ref{diseqn0} must be adapted to

\begin{equation}
\frac{1}{d^2}\left(\frac{d^2 \Omega_b N_{HI}}{\Delta V}\right) D^2 t^{1/2} > 
k_2
\end{equation}

\noindent where $N_{HI}$ is the column density in cm$^{-2}$ and $k_2$ is 
a constant.  As $\Omega_b = (\lambda/D)^2$

\begin{equation}
\frac{N_{HI}}{\Delta V} > \frac{k}{\sqrt{t}}
\label{diseqn2}
\end{equation}

\noindent where $k$ is a constant. 
Equation \ref{diseqn2} is independent of $D$, dish diameter, and is a mandatory
requirement for detection because a source which cannot be detected when it
fills the beam certainly cannot be detected when it does not.  In other
words, short integration-time surveys are only sensitive to high
column-density sources, {\it irrespective of dish size}, a limitation which
is seldom acknowledged (e.g. Zwaan et al. 1997). (See Appendix 
\ref{col-density-sens} for a full derivation of Equation \ref{diseqn2})

We now show that the volume in which a galaxy can be detected only
depends on its peak flux, and is independent of the actual column
density, as long as the peak flux is higher than the survey limit.

First let us consider the case of a galaxy with mass $M_{HI}$, and
velocity width $\Delta V$, which just fills the beam (case (a) in
Fig.~1).  We will also assume that it is at the peak flux limit $S$ of
the survey.  As the latter limit only depends on the flux per velocity
channel we find that for this galaxy 

\begin{equation}
M_{HI} \propto S (\Delta V) d^2 \propto N_{HI} \, r_{gal}^2 \propto N_{HI} 
\, d^2 \theta^2
\end{equation}

\noindent or 

\begin{equation} S \cdot \Delta V \propto N_{HI} \Delta\Omega_{gal}
\end{equation}

\noindent where in this
case $\Delta\Omega_{gal} = \Delta\Omega_{beam}$.  As our survey is
peak-flux limited and $\Delta\Omega_{beam}$ is constant, we find that,
at fixed $\Delta V$, the limiting column density is directly
proportional to the limiting peak flux. In other words, a galaxy with a 
column density lower than the limiting
column density will also have a peak flux lower than the limiting
peak flux, and can thus never be detected.

\begin{figure}
\includegraphics[width=84mm]{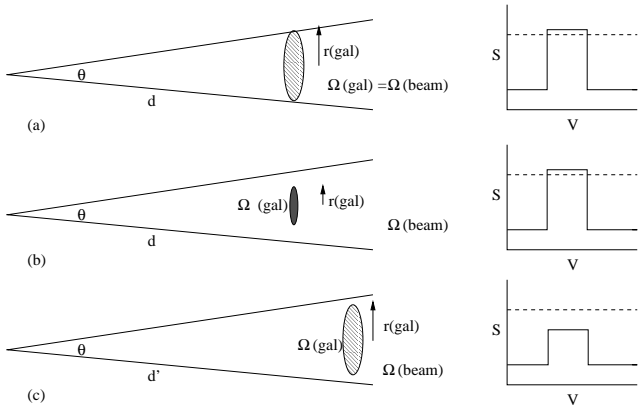}
\caption{Detectability of galaxies with different column densities in a survey.
We consider the case of a galaxy with mass $M_{HI}$ and velocity width
$\Delta V$. Panel (a): the galaxy just fills the beam ($\Omega_{beam}
= \Omega_{gal}$) and has a peak flux $S$ close to the survey limit (dashed
line in the cartoon spectrum on the right hand side). This galaxy will
have the lowest detectable column density. In panel (b) the galaxy is
smaller, does not fill the beam and has a higher column density. The
higher column density compensates for the beam dilution, and the
galaxy will still be detected, as shown on the right hand side. Panel
(c): as (a) except that the distance $d'>d$. The galaxy does not fill
the beam, but the low column density yields a peak flux $S$ lower than
the survey limit. This galaxy will not be detected.}
\label{nhi_dist}
\end{figure}

A more compact galaxy with the same $(M_{HI}, \Delta V)$ at the same
distance must necessarily have a higher column density (case (b)
in Fig.~1). This increase in column density compensates for the beam
dilution factor $r_{gal}^2/r_{beam}^2 =
\Delta\Omega_{gal}/\Delta\Omega_{beam}$, and the galaxy will still be
detected, as its peak flux $S$ is higher than the survey limit.

In contrast, if we take the galaxy from our original example (which
just fills the beam at distance $d$) and put it at a larger distance
$d'$ (case (c) in Fig.~1), the peak flux drops by a factor $(d'/d)^2$ 
and will not meet the survey limit.

The detectability of a galaxy is thus independent of its column
density, provided that its peak flux is higher than the survey limit.
A galaxy filling the beam at the peak flux survey limit
will  have the lowest detectable column density.  At a fixed
$(M_{HI},\Delta V)$, galaxies with higher column densities 
(necessarily) do not fill the beam, but will be detected over the same
volume, as they will exhibit identical peak fluxes.  Similar galaxies at
larger distances will not be detected, independent of column density, as
they will drop below the peak flux limit.

\begin{table}
\caption{Volume sampled for galaxies of different masses and column-densities
in Mpc$^3$, for a constant velocity width of 200 km\,s$^{-1}$}
\label{mhi-nhi_tab}
\begin{tabular}{lcccc}
&$10^7 M_\odot$&$10^8 M_\odot$&$10^9 M_\odot$&$10^{10} M_\odot$\\
$10^{18}$ cm$^{-2}$&0&0&0&0\\
$10^{19}$ cm$^{-2}$&0.22&6.9&220&6900\\
$10^{20}$ cm$^{-2}$&0.22&6.9&220&6900\\
$10^{21}$ cm$^{-2}$&0.22&6.9&220&6900
\end{tabular}
\end{table}

Table \ref{mhi-nhi_tab} demonstrates this for galaxies of
different masses and column-densities.  The $10^{18}$ cm$^{-2}$ galaxy is
below the column-density limit and will never be detected in HIDEEP.  All
the other galaxies are above the column-density limit and are therefore
detected out to the distances set by their peak fluxes.  The volume over which
a galaxy above the column-density limit will be detected is determined solely 
by the peak flux of the galaxy.

In summary, 21-cm surveys have two constraints: (a) a peak-flux limit given by
Equation \ref{diseqn0} in which dish size ($D$) is a distinct advantage, and 
(b) a surface-density limit where dish size is 
irrelevant, but in which integration time per beam is all-important.  In a 
search for high surface-density $N_{HI}$ objects, short integration-times per 
beam are favoured; in a search for low column-density (and therefore
low surface-brightness) 
objects, long integrations are mandatory.  (An alternative way of looking at 
this is to note that larger dishes project the same system noise on to smaller 
areas of sky, because their diffraction-limited beams are smaller, and so have
to contend with higher apparent sky-noise).  Thus HIDEEP and HIPASS are
complementary: HIPASS with its relatively short integration-time per beam 
(450s) picks up large numbers of high surface-density sources but is 
insensitive to objects below $1.6\times 10^{19}$ cm$^{-2}$ ($\Delta V = 200$ 
km\,s$^{-1}$) while HIDEEP with its very long integration-time per beam 
(9000s) is the first blind survey (see Table \ref{hitab}) capable of reaching
 much lower surface-density $N_{HI}$, and hence surface-brightness, limits

Before the advent of the multibeam 
system at Parkes it was neither profitable nor practical to make the very long 
integrations required to reach low column density.  Thus the limits quoted
in such surveys for the total amount of cosmic H{\sc i} referred only to 
high column-density clouds and galaxies, though this was rarely acknowledged.
Their integration times per beam were mostly far too short to detect the lower
surface-density features.  HIDEEP (see Appendix \ref{col-density-sens}) 
should and indeed does reach 

\begin{equation}
N_{HI} \geq 2.1 \times 10^{16} \Delta V
\label{diseqn3}
\end{equation}

\noindent $\simeq 4 \times 10^{18}$ cm$^{-2}$ for a typical galaxy with a
velocity width of 200 km\,s$^{-1}$.

In principle, therefore, we should be able to reach galaxies with lower 
surface-brightnesses than any detectable before either in H{\sc i} or
in the optical (see Table 1).

The H{\sc i} survey is complemented by
deep optical observations, reaching an isophotal level of 
26.5 mag\,arcsec$^{-2}$ in the {\it R}-band, covering three-quarters of the 
survey region.  For our analysis, we assume a value of $H_0 = 
75$~km\,s$^{-1}$\,Mpc$^{-1}$ throughout.

\section{Previous 21-cm Blind Surveys}
\label{previous-surveys}

Despite technical difficulties, blind surveys have been carried out, often 
using special techniques (see Table \ref{hitab} for details). For instance 
Shostak (1977) re-examined the signals in the `off-beams' of an NRAO 300-foot 
survey, where the `on-beams' were pointed at bright, optically selected 
galaxies. Latterly, Zwaan et al. (1997) and Schneider, Spitzak, \& Rosenberg 
(1998) have used the Arecibo radio telescope to carry out deep 
surveys. The results of such blind surveys have generally proved negative in 
the sense that very few previously-uncatalogued galaxies or intergalactic 
gas clouds (IGCs) were detected. 

\begin{table*}
\begin{minipage}{126mm}
\caption{Blind H{\sc i} sureys}
\label{hitab}
\begin{tabular}{lcccc}
&AHISS\footnote{Zwaan et al. 1997}&Arecibo Slice\footnote{Schneider et al. 
1998}&Shostak (1977)&Henning (1995)\\
\hline
Telescope&Arecibo&Arecibo&NRAO 300 ft.&NRAO 300 ft.\\
Channel separation (km\,s$^{-1}$)&16&16&11&22\\
Velocity range (km\,s$^{-1}$)&-700 -- 7,400&100 -- 8,340&-775 -- 
11,000&-400 -- 6,800\\
Noise channel$^{-1}$ beam$^{-1}$ (mJy)&0.75&2&18 -- 105&3.4\\
Ind. Sight-lines&6,000&14,130&6,050&7,200\\
FWHM&3.3\arcmin&3.3\arcmin&10.8\arcmin&10.8\arcmin\\
Area (deg$^2$)\footnote{Area within FWHM of beam}&13&33.6&154&183\\
$5\sigma M_{HI}$ limit ($M_\odot d_{\rm Mpc}^{-2}$)\footnote{For $\Delta V = 
200$ km\,s$^{-1}$, peak-flux limited}&
1.0$\times 10^{5}$&
2.8$\times 10^{5}$&
2.5 -- 15$\times 10^{6}$&
4.7$\times 10^{5}$\\
$5\sigma N_{HI}$ limit (cm$^{-2}$)$^d$&
1.8$\times 10^{19}$&
4.8$\times 10^{19}$&
4.1 -- 24$\times 10^{19}$&
7.7$\times 10^{18}$
\\
\\
&ADBS\footnote{Rosenberg \& Schneider 2000}&HIPASS&HIJASS (projected)&HIDEEP\\
\hline
Telescope&Arecibo&Parkes&Lovell&Parkes\\
Channel separation (km\,s$^{-1}$)&33.8&13.2&13.2&13.2\\
Velocity range (km\,s$^{-1}$)&-654 -- 7,977&-1280 -- 12,700&-3000 -- 10,000&
-1280 -- 12,700 \\
Noise channel$^{-1}$ beam$^{-1}$ (mJy)&3 -- 4&14&14&3.2\\
Ind. Sight-lines&181,000&610,000&380,000&670\\
FWHM&3.3\arcmin&15\arcmin&12\arcmin&15\arcmin\\
Area (deg$^2$)$^c$&430&30,000&13,000&32\\
$5\sigma M_{HI}$ limit ($M_\odot d_{\rm Mpc}^{-2}$)$^d$&
4.2 -- 5.5$\times 10^{5}$&
1.9$\times 10^{6}$&
1.9$\times 10^{6}$&
4.4$\times 10^{5}$\\
$5\sigma N_{HI}$ limit (cm$^{-2}$)$^d$&
7.2 - 9.6$\times 10^{19}$&
1.6$\times 10^{19}$&
2.6$\times 10^{19}$&
3.7$\times 10^{18}$\\
\end{tabular}
\end{minipage}
\end{table*}

However, as shown in Table \ref{hitab}, such surveys did not have sufficient
sensitivity to low column density ($N_{HI}$) gas to actually detect such 
objects.Quite apart from the theoretical arguments leading to Equation 
\ref{diseqn2}, the column-density 
sensitivity of any survey can be estimated retrospectively from its 
sensitivity to unresolved sources because (see Appendix 
\ref{col-density-calc}):

\begin{equation}
N_{HI}^{min}(\Delta V) = 4.5\times 10^{20} \left(\frac{F_{HI}^{gal}}{\Delta 
V^{gal}\delta\theta^2}\right)^{min}\times \Delta V
\label{nhieqn2}
\end{equation}
\noindent where $F_{HI}^{gal}$ and $\Delta V^{gal}$ are the integrated flux 
(in Jy\,km\,s$^{-1}$) and the velocity width (in km\,s$^{-1}$) of the source 
in the survey with the lowest value of $F_{HI}/\Delta V$ and $\delta\theta$ 
is the source size in arcmin.  When considering large LSB galaxies which could
fill the beam, we set $\delta\theta$ equal to the beam FWHM to find the
sensitivity to such systems.  That the left hand side of Equation 
\ref{nhieqn2}
is independent of dish size can be seen from the right hand side as both the 
minimum flux (top) and the beam size (bottom) are inversely proportional to the
dish area.  An examination of the various survey 
source-limits shows excellent agreement between the column density as 
calculated according to Appendix \ref{col-density-calc} 
and separately from applying Equation 
\ref{nhieqn2} to the various source lists. For instance if we look at the
deep drift-scan survey carried out by Zwaan et al. (1997), the
Arecibo H{\sc i} Strip Survey (AHISS), and examine those sources which lie 
within its main beam, we can use these to calculate the column-density 
limit of this survey for a typical velocity width of 200 km\,s$^{-1}$.  We
find that this limit is:

\begin{equation}
N_{HI}\hbox{(AHISS)} \geq 3.5\times 10^{19}
\label{ahiss-sens}
\end{equation}
\noindent i.e. $N_{HI}^{lim}$ (AHISS) $\simeq 10^{19.6}$ cm$^{-2}$ -- almost 
precisely
the minimum value found by the VLA follow-up observations.  In other words, 
the AHISS survey was a shallow survey capable only of 
picking up sources down to a few times $10^{19}$~cm$^{-2}$ with a 
velocity-width of $\simeq 200$~km\,s$^{-1}$.  The AHISS results do not, 
therefore, rule out the presence of a population of low column-density
galaxies.  It should be noted that the 5$\sigma$ column-density sensitivity
for AHISS given in Table \ref{hitab} is substantially lower than found in 
Equation \ref{ahiss-sens}.  This is consistent with the finding of Schneider
et al. (1998) that the limit for AHISS is well above 5$\sigma$.

The other recent deep Arecibo survey, the Arecibo Slice (Schneider et al. 
1998; Spitzak \& Schneider 1998) 
operated rather differently. Patches of sky about 2 beam diameters apart 
were followed for 60s, and any sources tentatively picked up were then 
re-scanned using a grid search. That meant that the sensitivity varied by a 
factor of $\simeq 4$ over the total 55 square degrees searched.  The median
flux of the sources detected over the velocity range 100 to 8340~km\,s$^{-1}$ 
was 2.69~Jy\,km\,s$^{-1}$ as against the median value for HIDEEP of 
1.96~Jy\,km\,s$^{-1}$. Because of the short 60s integration time, the 
sensitivity to low-column-density sources was poorer than AHISS. Nevertheless 
there were interesting results.  Only half the galaxies detected were in 
any optical catalogue  and about a third were LSB galaxies.  The 
galaxies have much lower bulge-to-disc ratios than found in optically-selected
samples.  The median $M_{HI}/L_B$ of the survey was $0.89 M_\odot/L_\odot$ 
and the galaxies with larger $M_{HI}/L_B$ ratios had lower 
surface-brightnesses.

Henning's (1992; 1995) survey with the Green Bank 300-foot telescope was 
largely behind the galactic plane and therefore not comparable with HIDEEP.  
The Arecibo Dual Beam Survey (ADBS; Rosenberg \& Schneider 2000; 2002) 
covered a larger area of sky than the
other surveys but to a much shallower depth, and so could not set any limits to
the population of low-column-density galaxies.

Although blind 21-cm surveys are, in principle, the ideal way of 
circumventing optical selection effects and looking for LSB galaxies and 
IGCs, the weakness of the 21-cm signal and the
system noise make it very difficult to find sources
unless they have high column-densities. Only very long integrations  
are capable of reaching low column-density limits, and possibly
finding objects of lower surface-brightness than can 
be seen optically -- thus the interest of HIDEEP. We claim that the 
arguments summarised in Equation \ref{disney-banks} and Table 
\ref{nhi-sb_tab} demonstrate that we should, for the first time at 21-cm, 
be capable of locating such objects. Even so, the limitations of all such 
blind H{\sc i} surveys should constantly be kept in mind.  Such surveys 
have lower sensitivity to broader-line sources (see below) and may, 
particularly at low column-densities ($< 10^{19}$~cm$^{-2}$), be severely 
affected by ionisation and spin temperature effects (Section 
\ref{col-densities}).  H{\sc i} surveys, 
therefore, can set only lower limits to the number of LSB galaxies and
IGCs in the cosmos.

\section{The HIDEEP survey}
\label{hideepsurvey}

\subsection{The H{\sc i} data}
\label{hi-sec}

HIDEEP was carried out in a region of Centaurus centred on 
$\alpha = 13^h40^m00^s$, $\delta = -30^\circ 00^\prime 00^{\prime\prime}$ 
(J2000) with 1024 spectral channels covering $-1280$ to 12,700 km\,s$^{-1}$. 
Due to the shape of the Parkes multibeam footprint, the survey has a
sensitivity as good as or better than HIPASS over 6 by 10 degrees with a 
uniform sensitivity over the central 4 by 8 degree area.  This region lies 
in the supergalactic plane, 30 degrees from the galactic plane.  The HIDEEP 
volume includes the Cen A group (Banks et al. 1999) and the outer parts of the 
Centaurus cluster.  The observations were carried out in the southern autumns
of 1997, 1998, 1999, and 2000.

The data were processed using the standard multibeam reduction techniques,
as described in detail by Barnes et al. (2001).  Continuum sources were
removed using {\sc luther} (Wright \& Stewart 2003, in preparation).
Once integrated, the data take the form of a cube with voxels (3D pixels) 4 by
4 arcminutes on a side and 13.2~km\,s$^{-1}$ deep.  The half-power beam width 
is 15 arcminutes and the data were smoothed in the velocity direction (to 
reduce ringing) as part of the reduction process, giving a velocity resolution 
of 18~km\,s$^{-1}$.  The data in adjacent voxels are therefore not entirely
independent.

The sky was Nyquist sampled 50 times and some 1800 separate samples
contributed to the signal in each voxel.  Median filtering of this large 
sample greatly reduces interference while the data were all 
taken at night to avoid solar radiation entering the beam sidelobes -- 
a major source of noise during the day.

\begin{figure}
\includegraphics[width=84mm]{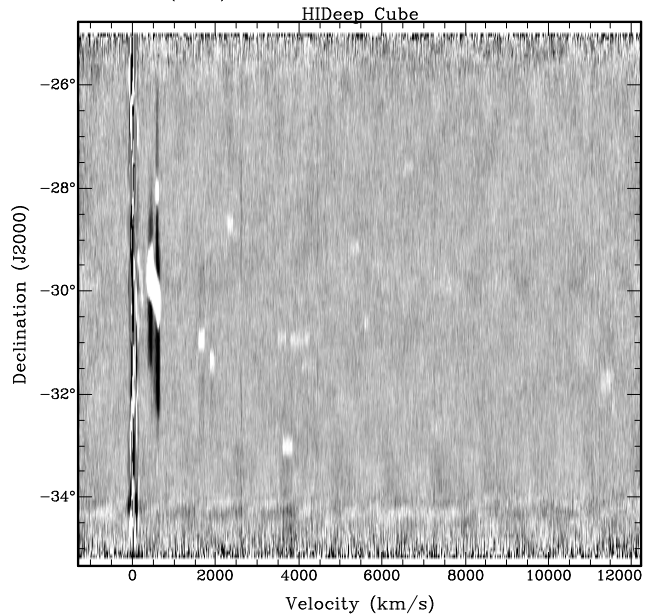}
\caption{A Decl. -- Velocity slice through the HIDEEP cube.  At least twelve 
sources can be seen, including Messier 83 at 500 km\,s$^{-1}$ and GSC 7265 
02190 at 11875 km\,s$^{-1}$ (Willmer et al. 1999)}
\label{cubepic}
\end{figure}

The final data cube can be examined in 3 planes: $(\alpha,\delta)$,
$(\delta, V)$ and $(V,\alpha)$, where $V$ is the
velocity direction.  All 3 planes are used 
for finding and measuring sources. Figure \ref{cubepic} shows a $(\delta,V)$
slice of the 
cube, showing the strong galactic signal at 0~km\,s$^{-1}$ as well as 12 
other sources including NGC 5236 (M 83) at 500~km\,s$^{-1}$ and GSC 7265 02190 
at 11875 km\,s$^{-1}$ (Willmer et al. 1999).  Continuum sources have been
removed and the 
nature of the remaining noise, against which sources must be found, can be 
seen.  The ripple seen just below -34$^\circ$ decl. is the residual of the 
strong continuum from the southern radio-lobe of IC~4296. The increased noise 
at the edges of the cube is the result of poorer sampling in these regions.

\begin{figure}
\includegraphics[width=84mm]{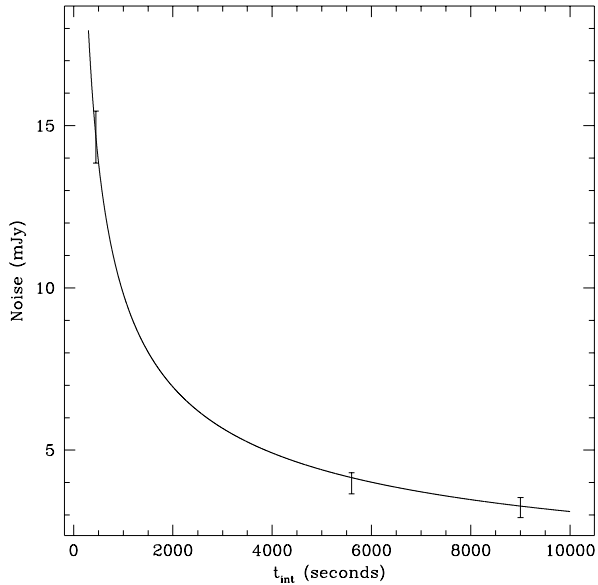}
\caption{Fall off of noise with integration time.  Points are for HIPASS, the
intermediate HIDEEP field with 5,600 seconds integration time, and the full 
HIDEEP field with 9,000 seconds integration time.  The line shows the 
theoretical $1/\sqrt{t_{int}}$ fall-off normalised to the noise in HIPASS.}
\label{deepnoise}
\end{figure}

Figure \ref{deepnoise} demonstrates the effectiveness of long
integrations.  We plot the median noise in
mJy\,beam$^{-1}$\,channel$^{-1}$ for integrations of
450~s\,beam$^{-1}$ (HIPASS), 5,600~s\,beam$^{-1}$ (Minchin 2001) and
9,000~s\,beam$^{-1}$ (HIDEEP).  As can be seen it falls to $3.2\pm 0.3$
mJy\,beam$^{-1}$\,channel$^{-1}$ against time in accordance with the
theoretical $1/\sqrt{t_{obs}}$. 

\begin{figure}
\includegraphics[width=84mm]{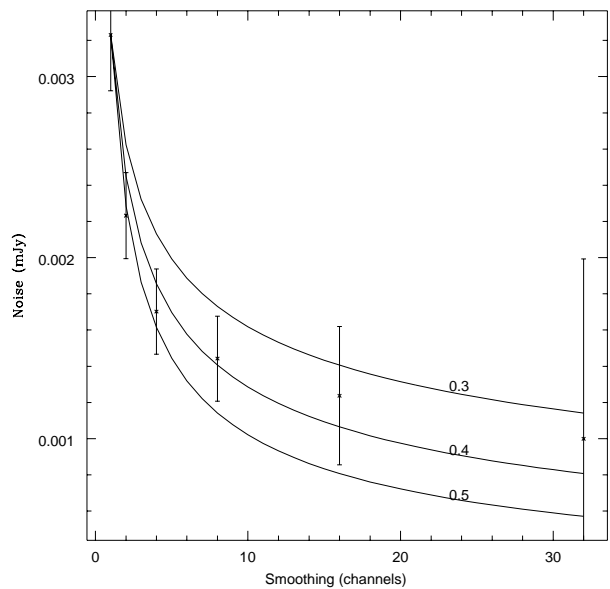}
\caption{Fall off of noise with smoothing.  The curves show $N^{0.3}$, 
$N^{0.4}$ and the theoretical curve from Poisson statistics, $N^{0.5}$ (where 
$N$ is the number of channels smoothed over).}
\label{smoothnoise}
\end{figure}

Figure \ref{smoothnoise} shows that smoothing in the velocity plane is
a much less effective way of reducing the noise.  We have smoothed the
data with a Hanning filter and removed every other channel from the smoothed
cube in order to leave only the independent channels. This has been
repeated to form cubes smoothed over 2, 4, 8, 16 and 32
channels. For white noise, the noise should fall as
$1/\sqrt{N_{chan}}$ (where $N_{chan}$ is the number of channels
smoothed over), however it can be seen that this is not the case for
$N_{chan} > 4$.  Beyond this the fall off is shallower than predicted
by Poisson statistics -- closer to $N_{chan}^{-0.3}$, although it is not
well described by a power-law.

\subsection{Source finding}
The HIDEEP cube was searched three times in an independent manner.
Two different people searched through the cube by eye, inspecting
every channel and noting down the sources found, and the third search
was carried out by an automated finding routine based on peak-flux
detection and template fitting.  This routine identified points higher
than 4.5-$\sigma$ on a Hanning-smoothed cube and fitted Gaussian templates 
with a large range of widths at these points, demanding a correlation of
better than 0.75.  The three lists given by these
searches were then compared and any sources found two or more times
were accepted.  The remaining disputed sources were examined by a third
member of the team for a final decision.  This team member had not previously
searched the data cube.

This gave us a final list of 173 sources, all of which have been judged
as real by at least two members of the team acting independently.
Accurate positions for all the sources were found by forming
zeroth-moment maps around their positions and velocities and
fitting Gaussians to these maps. This gives a positional accuracy, as
judged from those sources that can be securely identified with optical
galaxies, of around 2$^\prime$ (see Figure \ref{offsetcomp}).  The H{\sc i} 
parameters of the
galaxies were measured using the {\sc mbspect} routine in {\sc miriad} which
provides measures of the velocity width, the noise in
the spectrum and the peak flux of the source as well as robust
estimates of the integrated flux and systemic velocity. 



\subsection{Completeness of HIDEEP}

\begin{figure}
\includegraphics[width=84mm]{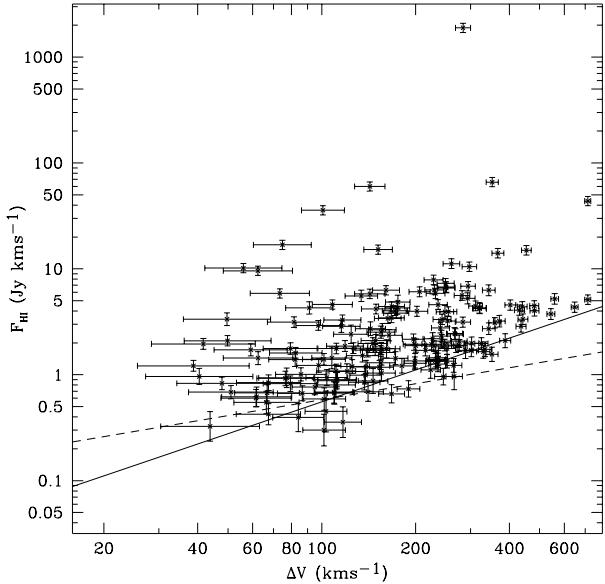}
\caption{Selection limits in velocity-width -- integrated flux space.  The 
theoretical $5\sigma$ limit for selection based on constant signal-to-noise 
using optimal smoothing is shown by the dashed line  and the $3\sigma$ limit 
for peak-flux selection ($F_{HI}/\Delta V_{20} = $ constant) is shown by the 
solid line.}
\label{selectionlimits}
\end{figure}

\begin{figure}
\includegraphics[width=84mm]{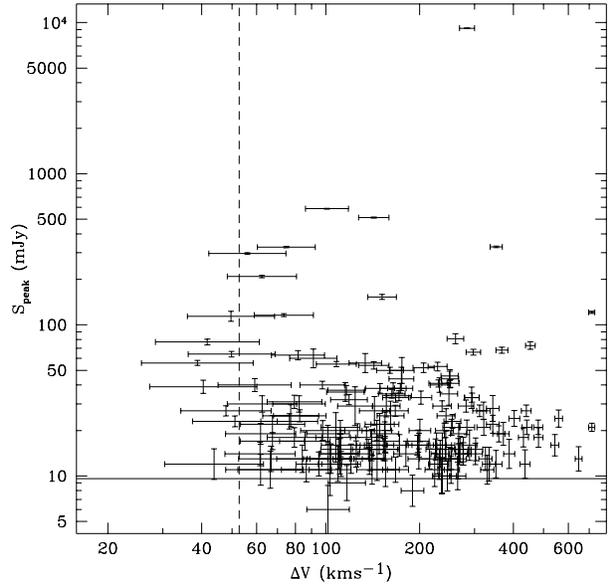}
\caption{Selection limits in velocity-width -- peak-flux space.  The $3\sigma$
(9.6 mJy) limit used above can be seen to be a good match to the selection 
limit of the data.  The dashed line shows $\Delta V_{20}$ = 4 channels 
(52.8~km\,s$^{-1}$).}
\label{selectionlimits2}
\end{figure}

The form of the selection present in the HIDEEP survey has been
analysed by plotting the integrated flux of the galaxies against their
velocity width (Figure \ref{selectionlimits}).  For a survey limited
solely by the total flux of the galaxies, the selection limit on this
graph would be a horizontal line.  If the best-possible real-world
selection was made, i.e. selection purely by signal-to-noise ratio using
optimal smoothing in the velocity plane, then the selection limit
would be a line with a 
slope of 0.5 on a log-log plot (assuming $SNR \propto 1/\sqrt{\Delta
  V}$, which is not strictly true as shown in Figure
\ref{smoothnoise}), shown here as a dashed line.  The solid line on
the graph shows a selection limit based on peak flux ($F_{HI} \propto
\Delta V_{20}$); only 16 of the 173 sources fall below this line and
only 8 of these are below by more than 1-$\sigma$.  Sources appear
to fall further below the line at the low-flux, low-velocity-width end, 
but this is probably due to the larger errors
in this region.  The peak-flux selection limit is shown in Figure
\ref{selectionlimits2}.  It can be seen that this explains well the
selection limit of the survey.

It can also be seen in both graphs that there is a paucity of galaxies
narrower than approximately 4 channel-widths (52.8~km\,s$^{-1}$).
This further selection effect, thought to be due to galaxies narrower than
this being indistinguishable from interference, is investigated in detail by 
Lang et al. (2003).

\begin{figure}
\includegraphics[width=84mm]{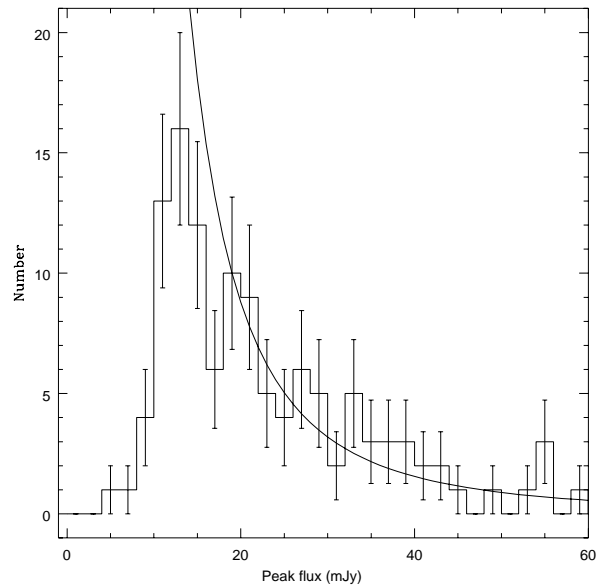}
\caption{Completeness of the HIDEEP survey: Source count against peak flux 
($S_{peak}$).  The histogram shows numbers found in each bin of peak flux, 
the curve represents $N(S_{peak})\propto S_{peak}^{-5/2}$ as expected for a 
flux-limited survey.}
\label{peaklim}
\end{figure}

\begin{figure}
\includegraphics[width=84mm]{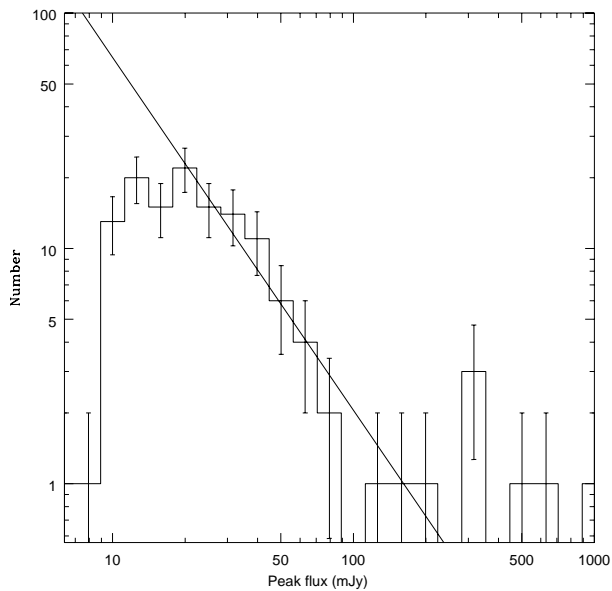}
\caption{Completeness of the HIDEEP survey: Log-log plot of source count 
against peak flux.  The solid line has a slope of $-3/2$, as expected for a 
flux-limited survey.}
\label{lgpeaklim}
\end{figure}

The completeness of the HIDEEP catalogue has been calculated
by looking at how the source counts vary with peak flux, as shown in
Figures \ref{peaklim} and \ref{lgpeaklim}.  For this analysis, 41 sources 
that may have
been detectable beyond the upper velocity limit were excluded in order to make
a purely flux-limited sample.  It can be seen from both figures that the
peak flux completeness limit is 18 mJy, or around 5.5$\sigma$.  This
is around the level where the completeness limit is expected to fall.
It can also be seen that large numbers of sources are found between
the completeness limit at 18 mJy and the selection limit of $\simeq
10$ mJy.

\subsection{The optical data}

To identify LSB galaxies it was necessary to obtain very deep optical data to
compare with the radio.  Accordingly eight 1-hour $6^\circ\times 6^\circ$
{\it R}-band Tech-Pan films 
were exposed at the UK Schmidt Telescope, centred on $13^h39^m50^s$~,$-30^\circ
00^\prime 12^{\prime\prime}$ (J2000).  These were digitised, linearised and 
stacked using the SuperCOSMOS machine at the UK Astronomical Technology Centre
in Edinburgh (Hambly et al. 2001).  The final image was then calibrated 
using the magnitudes of unsaturated ESO-LV (Lauberts \& Valentijn, 1989) 
galaxies within the region, which yields a calibration 
accuracy of approximately 0.2 magnitudes.  The Tech-Pan films used go
1 mag deeper than the IIIaF plates previously used at the UKST 
(Parker \& Malin 1999) and the digital stacking gives a further gain of 
over a magnitude compared to a single exposure
(Schartzenberg, Phillipps, \& Parker, 1996).  The limiting surface-brightness 
reached for small objects within the image is then 26.5~{\it 
R}\,mag\,arcsec$^{-2}$, equivalent to between 27 and 28
{\it B}\,mag\,arcsec$^{-2}$. 

Within the overlap region between the radio and optical images we find 96 H{\sc i} sources which have 
been uniquely identified with optical counterparts.

\begin{figure}
\includegraphics[width=84mm]{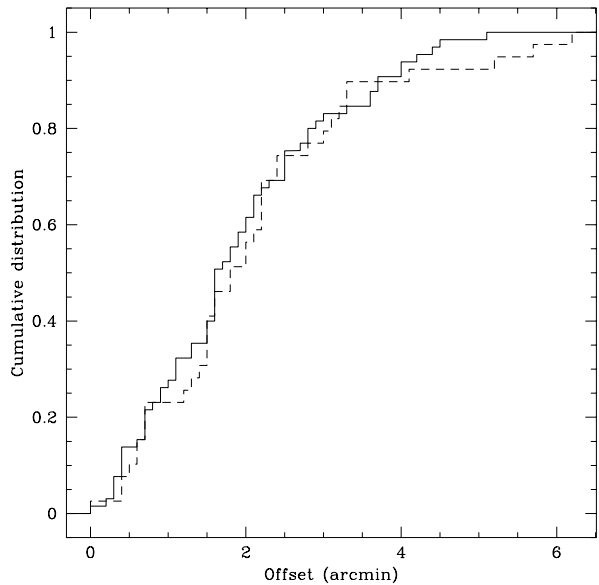}
\caption{Comparison between the cumulative distributions of offsets for the 
firm identifications (solid line) and the less certain counterparts (dashed 
line).}
\label{offsetcomp}
\end{figure}

Sources have been identified with galaxies on the optical image, firstly on 
the basis of positional coincidences.  Multibeam survey positions are 
generally accurate to 2 arcminutes (Koribalski et al. 2003), but this can be 
checked for the 65 
optical counterparts which have previously published optical velocities or 
for which we have our own optical spectroscopy or 21-cm interferometry data.  
A comparison can be made between the (radio -- optical) offsets of these firm 
detections with the remainder (Figure \ref{offsetcomp}).  A  Kolmogorov-Smirnov
test confirms that there is no significant difference between the 
distributions, implying that the purely positional coincidences can generally 
be trusted.  There may still be one or two incorrect identifications, but the 
tail of offsets out to 6.5 arcminutes probably reflects the 
positional accuracy of the HIDEEP survey.

In the overlap area 59\% of the sources are identified with previously
catalogued galaxies with matching redshifts, 24\% with previously 
catalogued galaxies without redshifts, and 17\% are previously uncatalogued
galaxies.  It appears that there are no 
intergalactic gas clouds unassociated with optical counterparts: all 
those galaxies which have not been uniquely associated with a counterpart have
more than one plausible candidate.

\begin{figure}
\includegraphics[width=84mm]{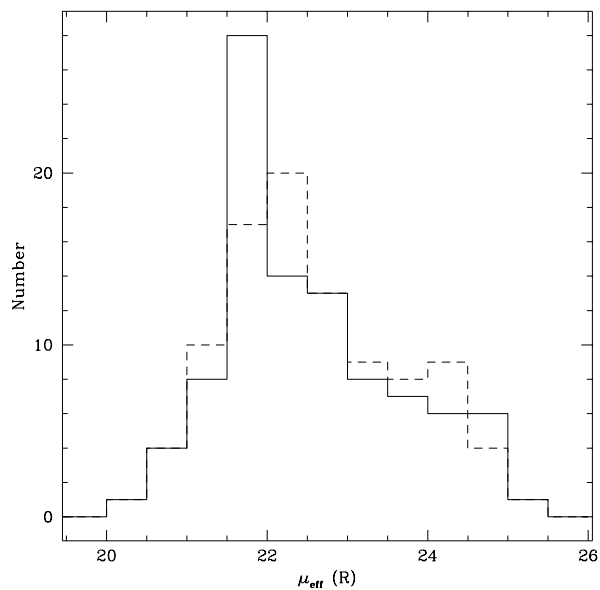}
\caption{Number of galaxies found in each surface-brightness bin.  The 
solid line shows the distribution of observed surface-brightnesses, the 
dashed line shows the distribution of surface-brightnesses after correction 
for galactic absorption, cosmological dimming, and inclination.}
\label{sbhist}
\end{figure}

We have measured effective radii and surface-brightnesses by fitting to the
surface-brightness profiles of the optical sources using the {\sc ellipse}
routine in IRAF.  The total magnitudes used were determined by {\sc 
sextractor} (Bertins \& Arnouts 1996).  The surface-brightness distribution
(Figure \ref{sbhist}) is  much broader than one finds in optically selected 
samples. A Kolmogorov-Smirnov test of our
distribution and the surface-brightness distribution of the ESO-LV shows 
that the hypothesis that both are drawn from the same parent population has 
a significance of less than 1\% -- this is due to the larger
number of LSB galaxies seen in the HIDEEP sample.  This confirms that H{\sc i}
surveys do, as expected, avoid some of the surface-brightness selection
effects present in optical surveys.  A fuller description and analysis 
of the optical properties will be given in a second paper (Minchin et al. 
2003, in preparation).

\subsection{Internal H{\sc i} correlations}

A correlation between H{\sc i} mass and observed velocity width of the form
$\Delta V \propto M_{HI}^\beta$ (where $\Delta V$ is the velocity width
uncorrected for inclination) is expected in the
HIDEEP data as it has been seen in optically-selected samples (e.g. Briggs
\& Rao 1993) and as it would be the consequence of the H{\sc i} 
Tully-Fisher relationship.

\begin{figure}
\includegraphics[width=84mm]{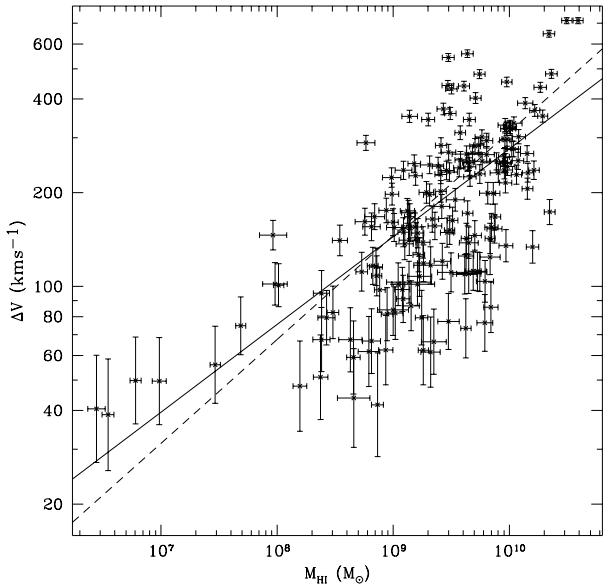}
\caption{H{\sc i} mass -- velocity width relationship for HIDEEP
  sources. The best fit to the HIDEEP data  
($\Delta V = 0.42^{+0.30}_{-0.17} M_{HI}^{0.282\pm 0.025}$) is shown by a 
solid line, while the  fit of $\Delta V = 0.15 M_{HI}^{1/3}$ (Briggs \& Rao 
1993) is shown as a dashed line.} 
\label{lmhi-ldv}
\end{figure}

That such a relationship can be seen in the HIDEEP data is shown in 
Figure \ref{lmhi-ldv}.  The solid lines shown here is for the best fit of 
$\Delta V_{20} = 0.42^{+0.30}_{-0.17} M_{HI}^{0.282\pm0.025}$ .  The dashed 
line shows $\Delta V = 0.15 M_{HI}^{1/3}$ as found 
by Briggs \& Rao (1993).  The best-fit slope found for the HIDEEP data 
is $2\sigma$ shallower than this, however this may be due to the selection
against narrow velocity-width galaxies seen earlier.  We cannot, therefore,
conclusively say that our sample shows a different relationship to that found 
by Briggs \& Rao.

It is also expected that there will be a relationship between peak flux and 
the value of $F_{HI}/\Delta V_{20}$ -- the peak flux of a top-hat function 
with a width of the 20\% width of the source and containing the same total 
flux.  This relationship is important for calculated H{\sc i} mass limits 
in a peak-flux limited survey, as it is necessary for relating the peak-flux 
limit to the integrated flux ($F_{HI}$) on which the H{\sc i} mass depends.

\begin{figure}
\includegraphics[width=84mm]{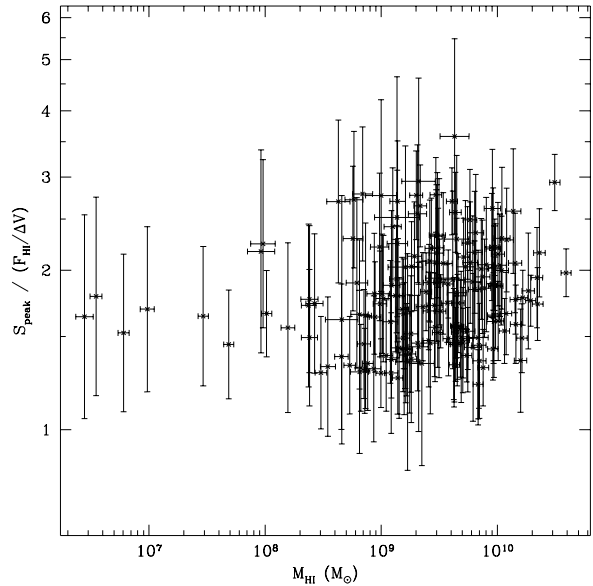}
\caption{Comparison of the ratio $\frac{S_{peak}}{F_{HI}/\Delta V}$ with 
$M_{HI}$.  The slope of the best-fit to this data is indistinguishable from 
zero, we 
therefore use the median value, $1.7\pm 0.3$ to describe the ratio across 
all H{\sc i} masses.}
\label{tophat-peak}
\end{figure}

However, as this relationship depends on the profile shape it may well vary 
with H{\sc i} mass.  This is investigated in Figure \ref{tophat-peak}.  This 
figure shows that there is no dependence of the ratio 
$\frac{S_{peak}}{F_{HI}/\Delta V}$ on $M_{HI}$: the slope of a fit to the 
points is statistically indistinguishable from zero.  We therefore take the
ratio to be a single number for all H{\sc i} masses, using the median value: 
$1.7\pm 0.3$.  This value was used to calculate the $3\sigma$ peak-flux 
limit in Figure \ref{selectionlimits}.

\begin{figure}
\includegraphics[width=84mm]{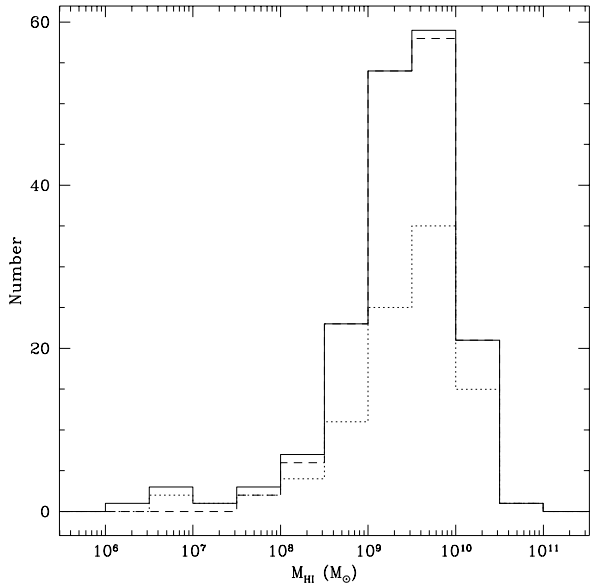}
\caption{Distribution of H{\sc i} masses.  The solid line shows the 
distribution for the whole survey, the dashed line shows the distribution 
excluding galaxies in the Centaurus A group and the dotted line shows the 
distribution of galaxies included in the optical sample.}
\label{mhidist}
\end{figure}

The H{\sc i} mass distribution of the HIDEEP sources is shown in Figure 
\ref{mhidist}.  This shows both the distribution of all the sources in HIDEEP 
(solid line), the distribution if the Centaurus A group galaxies are ignored 
(dashed line) and the distribution of galaxies included in the optical sample
(dotted line).
It can be seen that most of the low-mass galaxies are, unsurprisingly, in the 
Centaurus A group but we find only one high-mass galaxy (Messier 83).
There are 22 
hydrogen giants with $M_{HI} > 10^{10} M_\odot$, but only NGC 5291 has 
a mass greater than $3\times 10^{10} M_\odot$.  There are no galaxies detected
outside the range $10^6 M_\odot < M_{HI} < 10^{11} M_\odot$.  The optical 
sample is similar in shape to the full sample, but contains fewer sources.  
This is mainly because it covers a smaller area.  Some sources have also 
been omitted as a unique optical counterpart could not be identified.

\subsection{Sensitivity and comparisons with previous surveys}

Using the completeness limits and relationships found in the H{\sc i}
data, we can calculate the sensitivity limit of the HIDEEP
survey to galaxies of different masses and different column-densities.  
The mass (in solar masses) is related to the integrated flux by 
the equation

\begin{equation}
M_{HI} = 2.356 \times 10^5 F_{HI} d_{Mpc}^2
\label{himasseqn}
\end{equation}

\noindent (for $F_{HI}$ in Jy\,km\,s$^{-1}$) and the total flux is related to
the peak flux and the velocity width (Figure \ref{tophat-peak}) by

\begin{equation}
S_{peak} = (1.7 \pm 0.3) \times \frac{F_{HI}}{\Delta V_{20}}
\end{equation}

This allows the mass to be related to the peak flux as:

\begin{equation}
M_{HI} \simeq 1.386 \times 10^5 S_{peak} \Delta V_{20} d_{Mpc}^2
\end{equation}

However the velocity width is a function of the mass, 
$\Delta V \simeq 0.15 M_{HI}^{1/3}$ (although with a large scatter). If
we use the relationship found by Briggs \& Rao (1993) then

\begin{equation}
M_{HI} \simeq \left(0.15 \times 1.386 \times 10^5 S_{peak} 
d_{Mpc}^2\right)^{3/2}
\end{equation}

\noindent which gives:

\begin{equation}
M_{HI} \simeq 3.0 \times 10^6 S_{peak}^{3/2} d_{Mpc}^3
\end{equation}

\noindent and putting in the completeness limit of $S_{peak} = 0.018$~Jy gives:

\begin{equation}
M_{HI} \simeq 7.24 \times 10^3 d_{Mpc}^3
\label{mhi-sensitivity}
\end{equation}

\noindent which is the  sensitivity limit for sources of different masses
at different distances. Since the relationship between 
velocity-width and mass has a large scatter this is not the absolute 
detection limit: sources narrower than predicted by this relationship will
be seen further out, and sources wider than predicted will be found over a 
smaller volume.  However, it does indicate the distance to 
which most sources of a given mass will be seen.

The relationship between mass and velocity-width 
suggests the possibility of a selection effect in H{\sc i} surveys that 
appears to have been neglected in many previous surveys: that of a minimum 
believable velocity width.  Below a certain mass, the velocity-widths of many 
of the galaxies will be smaller than the minimum believable velocity-width of 
the survey -- the width at which a peak in the data is recognised to be a 
galaxy rather than a noise peak -- and will not be catalogued.  A fuller
analysis of this effect, using data from the H{\sc i} Jodrell All Sky
Survey, is given in Lang et al. (2003).

Examination of the HIDEEP data shows the minimum believable velocity-width
to be around 4 
channels wide for $\Delta V_{20}$, or 52.8 km\,s$^{-1}$.  This selection 
effect means that detecting large numbers of low-mass galaxies will require 
not only sensitivity but also narrow channel-widths.  However, this effect
could remove at most $\sim 40$ per cent of $10^7 M_\odot$ galaxies here.  
If the thermal broadening of the H{\sc i} is also taken into account, this 
percentage will fall.  It is therefore
unlikely that this will significantly change the shape of H{\sc i} mass 
functions down to
their current mass limits.  For HIDEEP, the distance at which galaxies that 
would be above the mass limit will fall below the minimum believable 
velocity-width limit is about 18 Mpc.

\begin{figure}
\includegraphics[width=84mm]{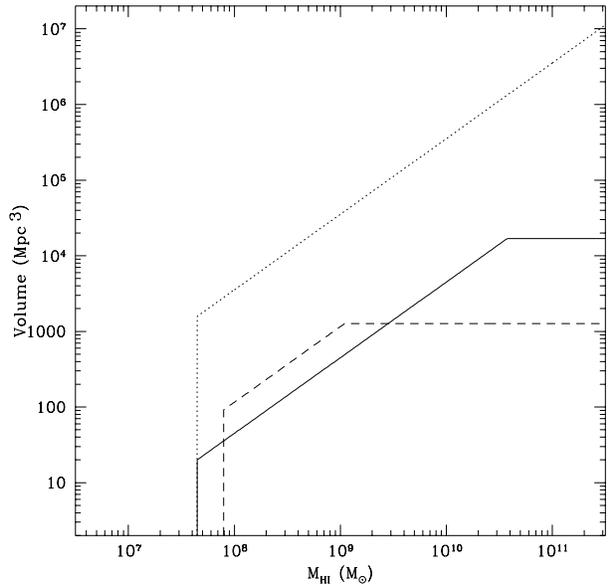}
\caption{H{\sc i} mass limits for three surveys:  HIDEEP (solid line), the 
HIPASS Bright Galaxy Catalogue (BGC; dotted line) and the Arecibo H{\sc i} 
Strip Survey (AHISS; dashed line).}
\label{mhilimits}
\end{figure}

Figure \ref{mhilimits} compares the H{\sc i} mass sensitivity of three 
surveys: HIDEEP, the HIPASS Bright Galaxy Catalogue (Koribalski 2003; BGC) 
and the Arecibo H{\sc i} Strip Survey (AHISS).  A minimum believable 
velocity-width of 4 channels has been applied to all these surveys although
galaxies narrower than this can be seen if they have high peak signal to 
noise ratios.
This will apply particularly to the HIPASS BGC, where the peak-flux cutoff 
of 116 mJy is more than 8$\sigma$.  For AHISS a limit of 5.25 mJy, or 
seven times the stated noise value of 0.75 mJy per channel, was adopted.  
This appears consistent with an analysis of the fluxes and velocity widths 
from Zwaan (2000) and with the analysis of Schneider et al. (1998) which 
concluded that the AHISS survey was limited at 7$\sigma$ due to the method 
of confirming sources.  Only the primary beam area was used in calculating 
the volume covered by AHISS.

Having estimated the sensitivity to sources in the mass-limited regime, we 
can also estimate the sensitivity to sources in the column-density limited 
regime.  Column-density sensitivity has often been presented, in a similar 
manner to optical surface-brightness, as not having a dependence on distance.
However the relationship between $\Delta V$ and $M_{HI}$ and the dependence 
of column-density sensitivity of $\Delta V$ (see Equation \ref{nhieqn2}) 
means that this is not the case for single-dish surveys -- the tendency
of higher mass sources to have larger velocity widths means that they will
(a) be seen to greater distances due to their higher mass and (b) will have
a higher column-density limit due to their larger velocity width.

The column-density of a source filling the beam (in $M_\odot$~pc$^{-2}$)
is given by:

\begin{equation}
N_{HI} = \frac{M_{HI}}{\pi\theta^2 d^2_{pc}} M_\odot \hbox{ pc}^{-2}
\end{equation}

\noindent where $\theta$ is the HWHM of the telescope beam in radians and 
$d_{pc}$ is the distance of the source in parsecs.  It can easily be seen 
that $N_{HI}$ is only independent of distance if the sensitivity of the 
telescope to $M_{HI}$ goes as $d^2$.  This would only be the case if galaxies 
at all different H{\sc i} masses had the same velocity-width or if the survey
was limited by integrated-flux -- neither of which are likely. 
There must, therefore, be a distance dependence for column-density 
sensitivity to average galaxies which follow the relationship between
$M_{HI}$ and $\Delta V$.

Putting in the $M_{HI}$ sensitivity from Equation \ref{mhi-sensitivity} and 
the beam size of Parkes (15 arcmin) gives a column-density sensitivity for 
HIDEEP of:

\begin{equation}
N_{HI} \simeq 6.1 \times 10^{16} d_{Mpc} \hbox{ cm}^{-2}
\end{equation}

\begin{figure}
\includegraphics[width=84mm]{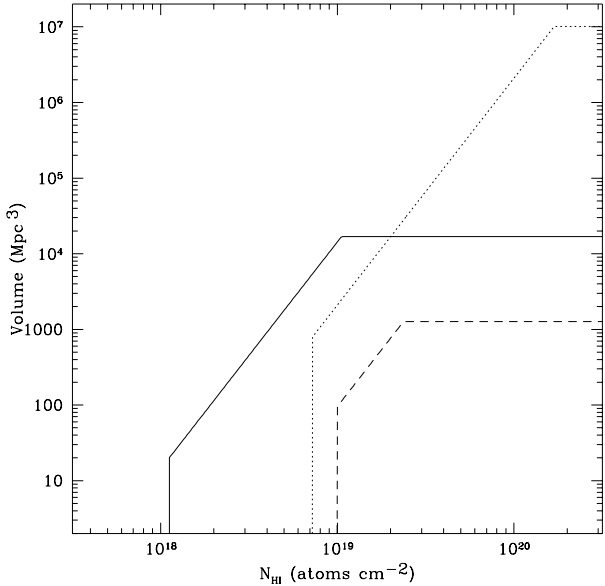}
\caption{H{\sc i} column-density limits for three surveys:  HIDEEP (solid 
line), the HIPASS Bright Galaxy Catalogue (BGC; dotted line) and the Arecibo 
H{\sc i} Strip Survey (AHISS; dashed line).  It can be seen that HIDEEP is 
considerably more sensitive than previous surveys to low column-density 
sources.}
\label{nhilimits}
\end{figure}

Again, this will be affected by the minimum believable velocity-width limit.  
Column-density limits for the three surveys (HIDEEP, HIPASS BGC and AHISS) 
are given in Figure \ref{nhilimits} where it can be seen that HIDEEP has a 
considerable advantage.

\begin{figure}
\includegraphics[width=84mm]{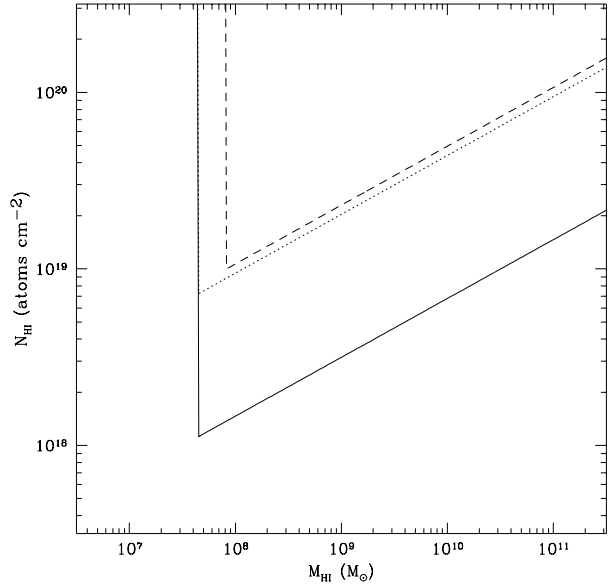}
\caption{Coverage of $M_{HI}$ -- $N_{HI}$ space for three surveys:  HIDEEP 
(solid line), the HIPASS Bright Galaxy Catalogue (BGC; dotted line) and the 
Arecibo H{\sc i} Strip Survey (AHISS; dashed line).  This shows that neither 
the HIPASS BGC nor the AHISS were likely to find low surface-brightness giant
galaxies.  However, if these exist and are not very rare, they should be 
found by HIDEEP.}
\label{mhi-nhi}
\end{figure}

Figure \ref{mhi-nhi} shows the region of $M_{HI}$ -- $N_{HI}$ space in which 
each survey is sensitive.  HIDEEP extends the region of $M_{HI}$ -- $N_{HI}$ 
space explored.  For instance, neither the HIPASS BGC or AHISS would find 
giant LSB galaxies unless they had very high $(M_{HI}/L_B)$s or very low
velocity widths.

\subsection{H{\sc i} properties}

The H{\sc i} properties of a sample of HIDEEP sources are given in Table {4}
(full table available online).  
Column 1 gives the HIDEEP identification for the
source.  Columns 2 and 3 give the right ascension and declination from
fitting to the zeroth-moment map.  Columns 4 -- 6 give the noise
as measured on the part of the spectrum not containing signal
($\sigma$), the integrated flux (zeroth-moment, $F_{HI}$), and
the peak flux ($S_{peak}$), all as measured by {\sc mbspect}.  Columns
7 and 8 give the systemic heliocentric velocity (first-moment,
$V_\odot$) and the velocity width at 20\% of the peak flux ($\Delta
V_{20}$), measured by {\sc mbspect} in the radio-velocity frame ($c
\frac{\Delta \nu}{\nu}$) and converted to $cz$.  Column 9 gives the
distance in Mpc calculated from the CMB rest-frame velocity of the
sources, which is approximately $V_{CMB} = V_\odot + 278\pm 8$
km\,s$^{-1}$ for the HIDEEP region (the exact correction varies across
the field).  This does not include any correction for bulk-motions
(Virgo-centric infall, etc.) nor does it include assigning sources
beyond the Centaurus A group to clusters and assigning a single
distance to that cluster.  Galaxies at less that 1000 km\,s$^{-1}$
have been assigned to the Centaurus A group at an assumed distance of 
3.5 Mpc. Column 10 gives the H{\sc i} mass calculated from $F_{HI}$ from 
column 5 and the distance from column 9 using the standard equation 
$M_{HI} = 2.356 \times 10^5 F_{HI} d^2_{Mpc} M_\odot$, where $F_{HI}$ is
in Jy~km\,s$^{-1}$.

\begin{table*}
\rotatebox{180}{\includegraphics{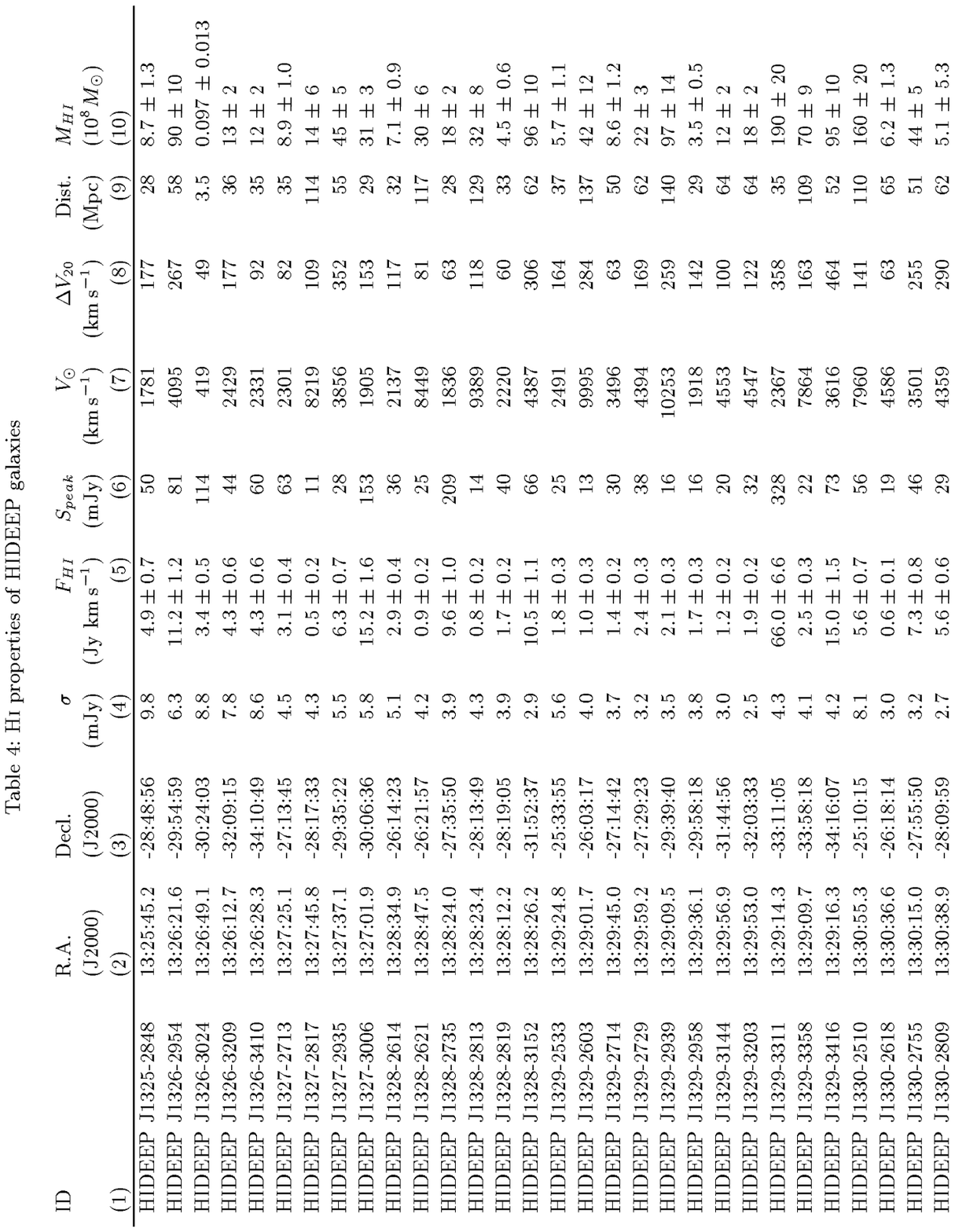}}
\end{table*}






\addtocounter{table}{1}

\subsection{Large-scale structure}

\begin{figure}
\includegraphics[width=84mm]{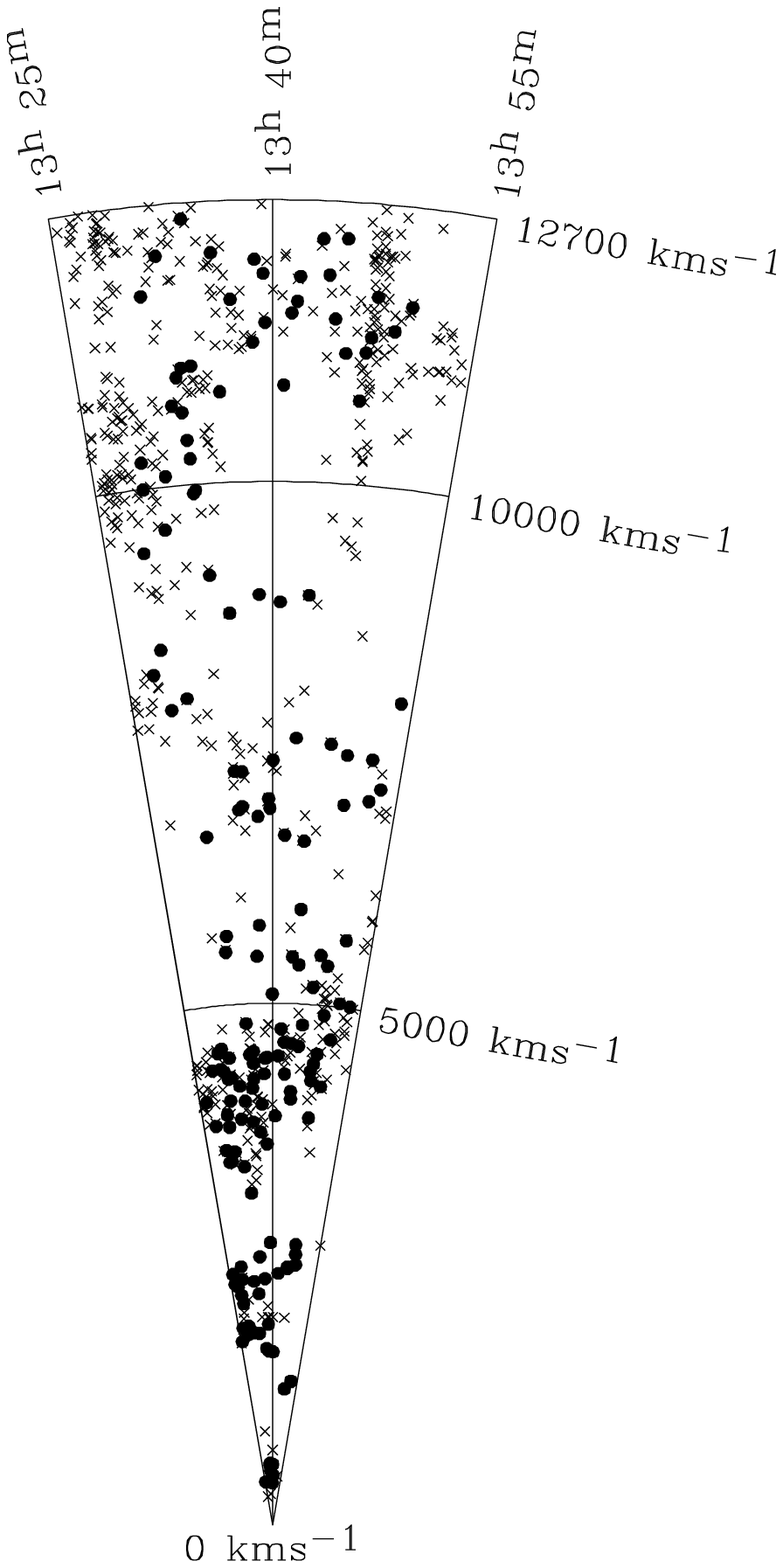}
\includegraphics[width=84mm]{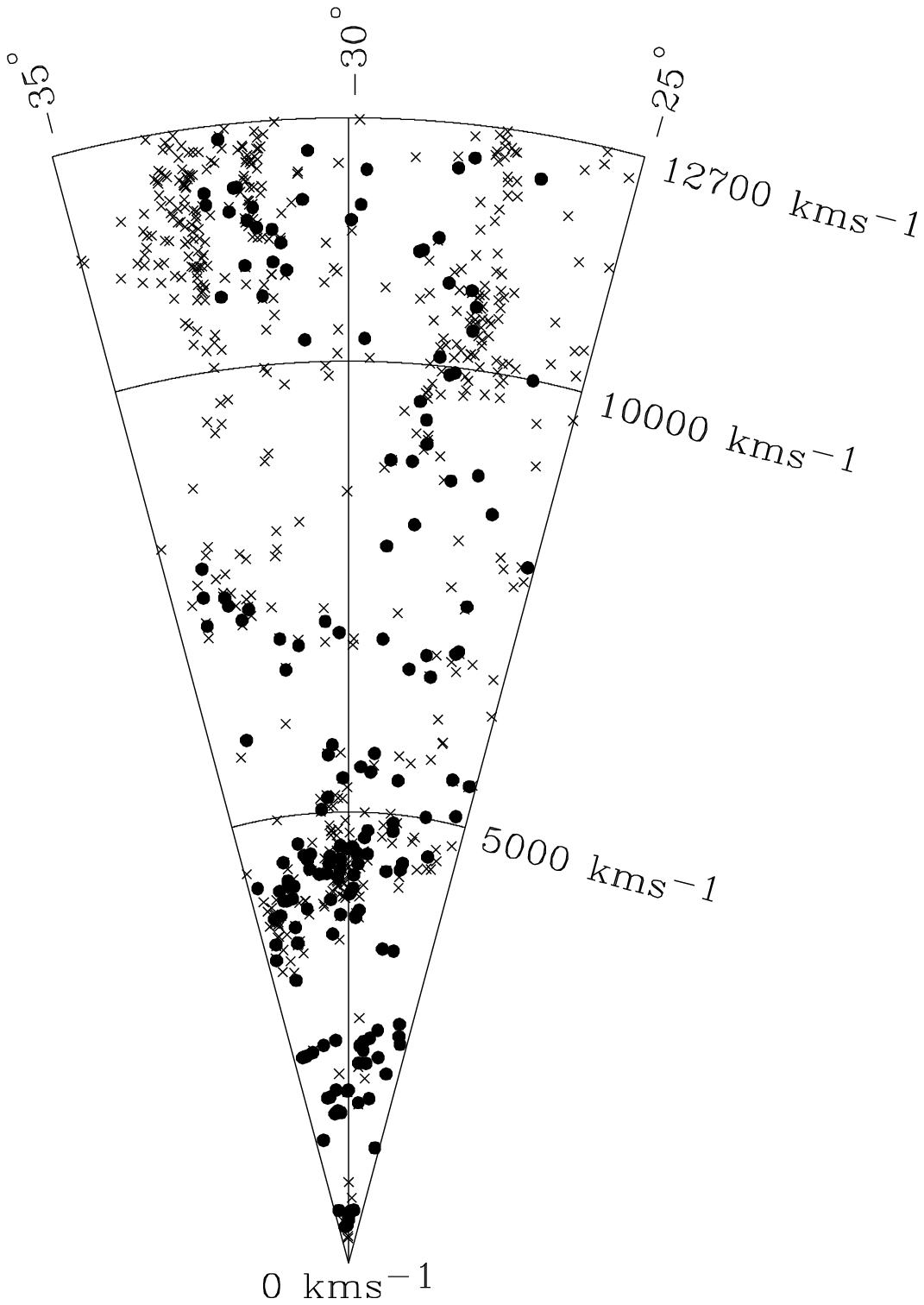}
\caption{Large scale structure of galaxies detected in HIDEEP (filled circles).
These pie-slices show the distribution in R.A. (upper) and Decl. (lower), 
with the angle expanded by a factor of three for clarity.  Galaxies from NED 
(crosses) are included for comparative purposes, it can be seen the H{\sc i} 
selected sample traces approximately the same large scale structure.}
\label{deeplss}
\end{figure}

The HIDEEP survey region lies in the supergalactic plane and there is,
therefore, much large scale structure in the survey volume.  This can
be seen in Figure \ref{deeplss}.  In particular, the Centaurus A
group, the Virgo Southern Extension and the Centaurus Cluster can be
seen.  Near the end of the bandpass a number of Abell clusters are
found, and more lie beyond the outer velocity edge of our survey.

\begin{figure}
\includegraphics[width=84mm]{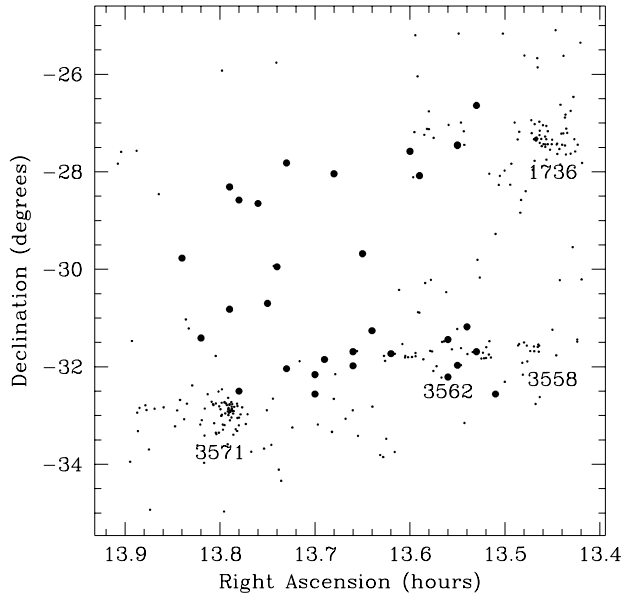}
\caption{H{\sc i} sources and galaxies in NED between 10,000~km\,s$^{-1}$ and 
the HIDEEP bandpass limit of 12,700~km\,s$^{-1}$.  The previously catalogued 
galaxies with redshifts are shown as dots and the HIDEEP sources are shown as 
filled circles.  Abell cluster numbers are superimposed on the concentrations 
of galaxies using positions from Abell, Corwin, \& Olowin (1989), these have
been shifted 0.75 degrees south for the sake of clarity.}  
\label{highvellss}
\end{figure}

Figure \ref{highvellss} shows that there is little correlation between
the distribution of previously catalogued galaxies with redshifts
above 10,000~km\,s$^{-1}$ and the distribution of the HIDEEP sources
in this velocity range.  Only the Abell 3562 cluster appears to have a
significant number of H{\sc i} detections and the H{\sc i}
sources also appear to populate the void at the north-west of the
region.  It is possible that the lack of correlation is due to the
targeting of optical redshift surveys towards the Abell clusters,
thus increasing the number of redshifts in those regions out of
proportion to their density.  While Abell 3571 and 1736 are both
centred inside the HIDEEP volume, at 11,500~km\,s$^{-1}$ and 
10,500~km\,s$^{-1}$ respectively, Abell 3558 and 3562 are both centred
beyond 14,000~km\,s$^{-1}$ (Abell, Corwin, \& Olowen 1989; Quintana \& de Souza
1993).  Only the low-velocity tails of these two clusters can be seen.

\subsection{Follow-up observations}

Follow up observations of some sources were carried out at 21-cm 
using the Australia Telescope Compact Array (ATCA) and optically using
the Double Beam Spectrograph on the ANU 2.3-m telescope at Siding
Spring Observatory.  The ATCA observations were carried out in the
375-m configuration in November 1999 and January 2000 and gave us
high-resolution 21-cm maps of the targeted sources in order to
accurately determine their positions.  Ten out of fourteen sources
were detected.  As the column-density sensitivity limit for these observations 
was around $10^{21}$~cm$^{-2}$, substantially higher than the limit for the
HIDEEP survey, the non-detections do not tell us anything about 
our survey limits.


Optical spectroscopy has
enabled us to positively identify ESO~509-G075 as the counterpart of
HIDEEP~J1335-2730 (which was undetected with ATCA) despite a previous
optical redshift (Quintana et al. 1995) placing it at twice the velocity of
the H{\sc i} source and allowed us to seperate
Abell~3558:[MGP94]4312 and 4317 which were too close together to be
separated by ATCA, identifying 4317 as the optical counterpart of
HIDEEP~J1334-3223.

\section{Inferred H{\sc i} Column-Densities}
\label{col-densities}

We have not yet obtained high-resolution H{\sc i} images of most of our 
sources so we do not know either their H{\sc i} radii or column densities.
We therefore calculate an inferred column-density ($N_{HI}^\circ$) from 
the H{\sc i} mass and the optical (effective) radius.  Because of our long 
integration time and low noise, we could in principle reach column densities
between one and two orders of magnitude deeper than previous blind surveys 
(Section \ref{previous-surveys}).  It is of interest to look for evidence, 
however indirect, as to whether this previously unexplored parameter space
is populated.

Salpeter and Hoffman (1996) found that $r_{HI} = (2.34 \pm 0.14) 
\times r_{B25}$ for an optically-selected sample of galaxies.  Obviously
$r_{B25}$ is not a good measure for the H{\sc i} radii of LSB galaxies,
which may not have a $B_{25}$ isophote.  However, the effective radius ($r_e$)
provides a model and surface-brightness independent measure of the optical 
size of galaxies.  In order to obtain a relationship between $r_e$ and 
$r_{HI}$, it is necessary to assume a relationship between $r_{B25}$ and
$r_e$.  This obviously introduces some unavoidable model-dependency into the 
analysis.  As the relationship between $r_{HI}$ and $r_{B25}$
is defined for HSB galaxies, we should define the relationship between
$r_e$ and $r_{B25}$ for a similar sample.  There is a further assumption
here that the proportionality between $r_e$ and $r_{HI}$ found
for these HSB galaxies will remain constant as we go to lower 
surface-brightnesses.  This may not be the case, and would introduce
systematic errors into our analysis, however $r_e$ is certainly a better
choice that $r_{B25}$ for looking at LSB galaxies.  We use the relationship
for disc galaxies in the ESO-LV catalogue (Lauberts \& Valentijn 1989)
that $r_{B25} = (2.15\pm 0.67) \times r_{e}^B$.  This then gives us:
\begin{equation}
r_{HI} =  (5.03 \pm 1.59) \times r_{e}^R
\label{rhieqn}
\end{equation}

\noindent From this we can calculate the inferred mean column-density 
$N_{HI}^\circ$ as:

\begin{equation}
N_{HI}^\circ = 10^{20.1} \frac{M_{HI}}{\pi R_{HI}^2} \hbox{ cm$^{-2}$}
\end{equation}

\noindent where $R_{HI}$ is the radius in pc, calculated from $r_{HI}$ using
the distance to the source.

\begin{figure}
\includegraphics[width=84mm]{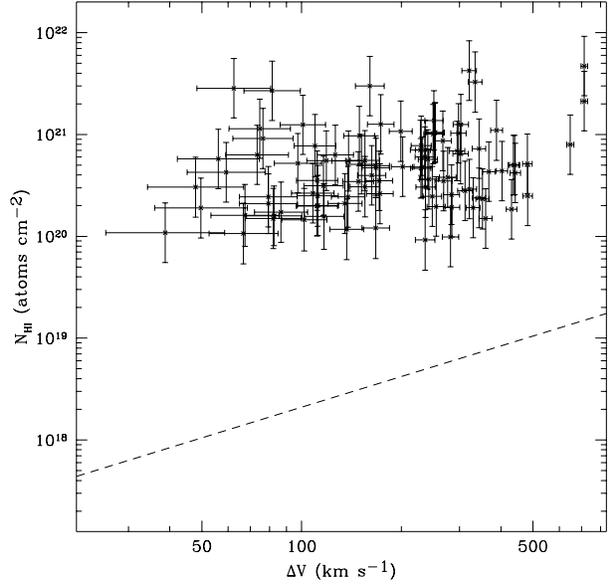}
\caption{Inferred column-densities of HIDEEP sources against 
velocity widths.  The limit shown is for the 18~mJy completeness limit for 
HIDEEP and sources filling the beam.}
\label{ldv-lnhi}
\end{figure}

Figure \ref{ldv-lnhi} shows these inferred column-densities plotted against
their velocity widths.  Above the survey limit (dashed line), the volume 
sampled depends only on H{\sc i} mass (or, more precisely, on peak flux) 
and not on column-density, as shown earlier in Figure \ref{nhi_dist}.
If low column-density galaxies were common, we would expect to see galaxies
at all column-densities in this plot.  However, this is clearly not the case 
-- there is a large gap between the lowest column-density galaxies found and 
our sensitivity limit.

\begin{figure}
\includegraphics[width=84mm]{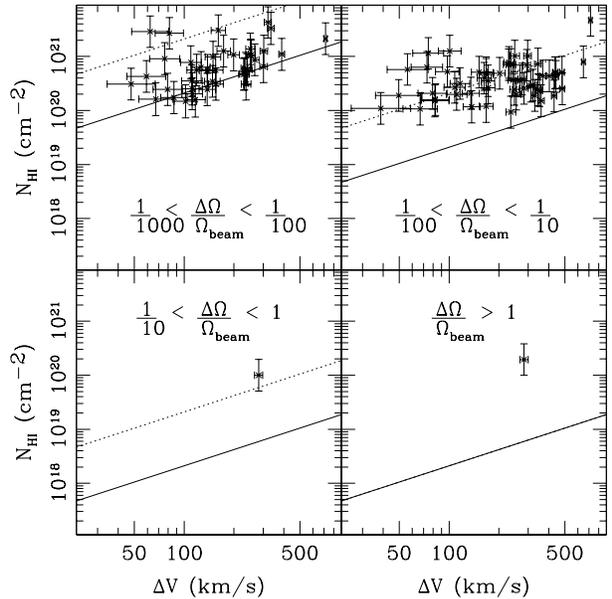}
\caption{Inferred column-densities of HIDEEP sources with different beam
filling factors against velocity widths.  The limits shown are for the
18~mJy HIDEEP completeness limit, the dotted line indicating the limit for
the lower filling factor and the solid line the limit for the higher
filling factor in each subgraph.}
\label{nhi_multi}
\end{figure}

That this gap is real and not an artefact of our method is shown in Figure 
\ref{nhi_multi}.  Here the sample has been split into four sections, according
to the beam-filling factor ($\Delta\Omega/\Omega_{beam}$) of the
sources, and the column-density limits calculated with this included.  
The top left panel shows that we are detecting galaxies up to the
column density limit of the survey \emph{for this range in beam
filling factors}. These are galaxies with small beam-filling factors,
and we can see that the inferred column density limits for these
galaxies are quite high (cf. Eqn B6).  The top right panel shows the
limits for galaxies with slightly larger beam filling
factors. Galaxies are detected with column densities close to the
survey limit \emph{for this range in beam filling factors}, but we can
see that they do not approach the limit as closely as in the previous
panel. This is an indication that in this sample galaxies with larger
beam filling factors (and hence lower column densities; cf.~Sect.~1)
are rarer.  This is confirmed by the two bottom panels. These show the
limits for galaxies that come close to filling the beam (and thus have the lowest column densities). We see that
despite the low limits, there are only two galaxies detected. This
thus indicates an absence of low-column density galaxies, despite
their potential detectability. 

Using the numbers of galaxies found within the overlap of the full-sensitivity
H{\sc i} survey and the optical survey, we calculate that, to 95\% 
confidence, low column-density galaxies make up less than 21\% of galaxies
with H{\sc i} masses between $10^8$ and $10^9 M_\odot$ (13 galaxies), less 
than 6\% between $10^9$ and $10^{10} M_\odot$ (52 galaxies), and less than 
19\% between $10^{10}$ and $10^{11} M_\odot$ (14 galaxies).


If our 
estimate of $R_{HI}$ were out by more than a factor of three, we would expect 
to see a lower limit to our column densities parallel to the plotted
completeness limit.  As the lower limit appears flat with respect to
velocity width, it appears that there are indeed no large very low 
column-density galaxies in our sample.  If such objects are present in the 
local Universe they must, therefore, be rare.  We discuss possible reasons for
this absence elsewhere (Disney \& Minchin 2003).

\begin{figure}
\includegraphics[width=84mm]{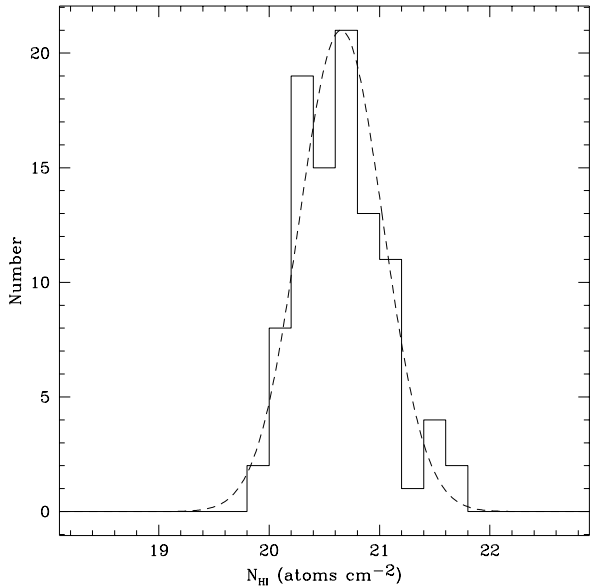}
\caption{Distribution of inferred column-densities of HIDEEP sources.
The dashed line indicates a Gaussian with a mean of 20.65 and a scatter 
of 0.38.  It can be seen that this is a fairly good description of the 
distribution.}
\label{nhihisto}
\end{figure}

\begin{figure}
\includegraphics[width=84mm]{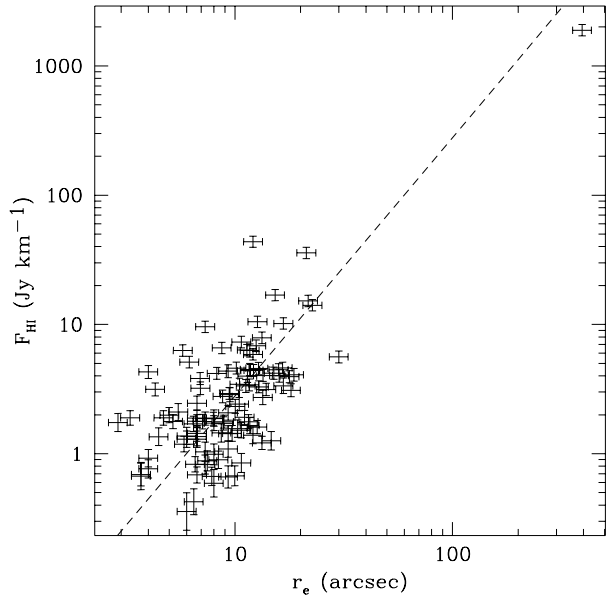}
\caption{Relationship between effective radius and H{\sc i} flux.  The 
dashed line shows the expected slope for $N_{HI}^\circ = 
10^{20.65}$~$cm^{-2}$.}
\label{reff-flux}
\end{figure}

The distribution of column densities appears to be the same at all 
surface brightnesses.  A two-dimensional Kolmogorov-Smirnov test (Peacock
1983) shows that the observed distribution is indeed consistent with a 
distribution having the same column-density distribution as HIDEEP (Fig. 
\ref{nhihisto}) at every 
surface brightness.  The distribution shown in Figure \ref{nhihisto},
is well described by a Gaussian with $\log(N_{HI}/\hbox{cm}^{-2}) = 20.65 
\pm 0.38$. Figure \ref{reff-flux} shows the measured data which gives rise to 
a constant column-density: a relationship between the effective radius
and the H{\sc i} flux.  The dashed line on this graph is for a constant
$N_{HI}^\circ$ of $10^{20.65}$ cm$^{-2}$, and it can be seen that this 
matches the data well.  This apparent constancy of column density is 
unexpected, we would expect to see a fall-off towards lower 
surface-brightness, as observed by de Blok et al. (1996) for a sample of 
LSB galaxies observed with the Westerbork Synthesis Radio Telescope (WSRT).

Swaters et al. (2002) observed a 
number of optically-selected galaxies with the WSRT, showing that the
relationship between H{\sc i} mass and diameter was much tighter when true
H{\sc i} diameters were used than when they were inferred from the
optical radii.  It is likely that much of the scatter in our relationship
has been similarly introduced by our use of optical radii. 
Observations of the true H{\sc i} diameters may well show a similar trend
to that seen by de Blok et al.

In order to compare our sample with the literature, we have taken the LSB
galaxy catalogue of Impey et al. (1996; ISIB96).  This survey provides 
effective surface-brightnesses and radii for all the galaxies found, across 
a wide range of surface brightness, and provides H{\sc i} mass measurements 
for a subsample of 190 galaxies.  We therefore calculate the column density 
from this data in exactly the same manner as for HIDEEP.  The Impey et al.
survey was carried out in {\it B}-band,
we therefore convert $\mu_e^B$ to $\mu_e^R$ using the average colour
for disc galaxies from de Jong 1996, $B - R = 1.1$.

\begin{figure}
\includegraphics[width=84mm]{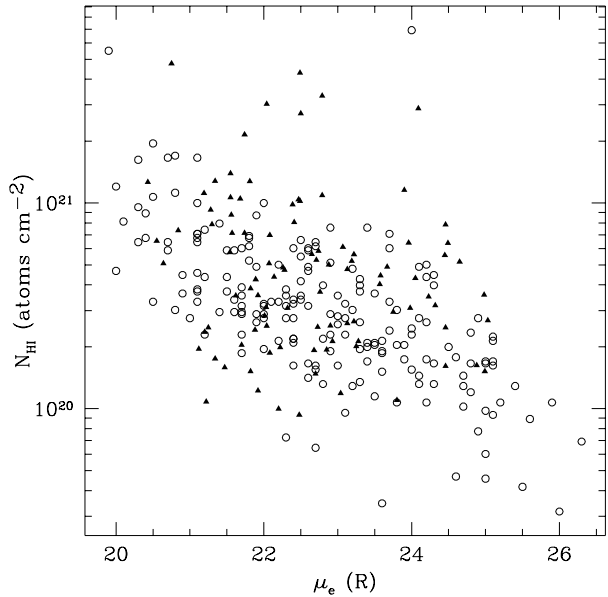}
\caption{Comparison between the surface-brightness -- column-density 
distribution for HIDEEP (triangles) and Impey et al. (1996), converted
to {\it R}-band assuming $B-R = 1.1$ (open circles).  Error bars have been
omitted for the sake of clarity.}
\label{sbcor-lnhi}
\end{figure}

Figure \ref{sbcor-lnhi} shows the surface-brightness -- column-density
distribution for both surveys.  HIDEEP galaxies are indicated by solid
triangles, ISIB96 galaxies by open circles.  It can be seen that 
there is a clear trend in the ISIB96 data, which is similar to that
seen by de Blok et al. (1996).  The surface-brightness distribution of
the ISIB96 galaxies is different to that of HIDEEP, due to the method in 
which the galaxies have been selected.  In order to 
compare the surveys we have therefore re-sampled the ISIB96 data to give
the same surface-brightness distribution (in 0.5 mag bins) as HIDEEP.  
This was carried out ten times, and the subsamples compared with HIDEEP
using the two-dimensional Kolmogorov-Smirnov test.  Half of the subsamples
were not significantly different at the 95\% level.

There are a number of systematics that could affect this comparison.
Our measure of $r_e$ could be systematically too small towards 
lower surface-brightnesses: this would lead to both $N_{HI}^\circ$ and $\mu_e$
being higher and so would act to destroy such a relationship between $N_{HI}$
and surface-brightness.  The  colours will not be exactly the same for all the
galaxies, introducing errors into the conversion of the  ISIB96 relationship to
{\it R}-band.  The low column-density galaxies ($N_{HI}^\circ < 
10^{20}$~cm$^{-2}$) in ISIB96 are generally large galaxies, with $r_{eff} > 
20^{\prime\prime}$.  Using our scaling between $r_{HI}$ and $r_{eff}$, this
gives them H{\sc i} sizes larger than the Arecibo beam.  It is therefore
likely that some of the hydrogen in these galaxies was missed in
the pointed observations, leading to lower inferred column-densities than 
is truly the case.  We cannot, therefore, say that the HIDEEP data is
inconsistent with ISIB96.  If there is a relationship between 
surface-brightness and $N_{HI}$ in our data, however, it is very weak indeed.

\section{Discussion, Conclusions and future work}
\label{conc-sec}

The existence of a large population of LSB galaxies could have  a significant
impact on several areas of astronomy, in particular on measurements of the 
luminosity and mass density of galaxies and on theories of galaxy formation 
and evolution. However, LSB objects are difficult to detect in the optical 
since, by definition, they are hard to discriminate from the background sky. 
Hence, alternative techniques need to  be considered for determining whether 
there exists  a significant population of extragalactic objects with optically
LSB. Searches for galaxies using the 21-cm line of atomic hydrogen provide one
such alternative method. If LSB galaxies have similar hydrogen content to 
galaxies of higher surface brightness, then we might expect that a 21-cm 
survey will not be biased against LSB galaxies. 

However, the detection of a galaxy in a 21-cm survey is a function of both 
its H{\sc i} column density and its H{\sc i} mass. If a galaxy has a column 
density below the column density limit of the survey then it will never be 
detected whatever its total H{\sc i} mass. Assuming LSB galaxies have similar 
H{\sc i} content to brighter galaxies but that this is distributed over a 
larger surface area, then we would expect LSB objects to have lower H{\sc i}
column densities.  Previous H{\sc i} surveys have not generally had long 
enough integration times to search to sufficiently low column densities to 
draw definite conclusions about the cosmic prevalence of gas-rich LSB galaxies.
HIDEEP with its 9000s\,beam$^{-1}$ integration time is the first survey with 
sufficient column density sensitivity to be able to place interesting limits 
on the existence and size of a low column-density/LSB population. Two major 
conclusions from HIDEEP have been presented here. 

Firstly, all of the sources found in HIDEEP appear to be associated with an 
optical counterpart on our deep UK Schmidt R-band data. In other words, we 
have not found an intergalactic H{\sc i} cloud down to our observational 
column density limit of N$_{HI}$/$\Delta$V = 2.1$\times$10$^{16}$ cm$^{-2}$ 
which corresponds to an N$_{HI}$ of 4.2$\times$10$^{18}$ cm$^{-2}$ for a 
typical galaxy velocity-distribution ($\Delta$V=200~km\,s$^{-1}$) and 
3.2$\times$10$^{17}$~cm$^{-2}$ for a typical QSOAL dispersion 
(15~km\,s$^{-1}$). Wherever neutral hydrogen is found it is accompanied by a 
visible population of stars above our surface-brightness limit of 26.5~R 
magnitudes per square arcsecond ($\sim$27.5 in B). 

Secondly, if we infer the H{\sc i} sizes of the HIDEEP detections from their 
optical sizes, then we can derive H{\sc i} column densities for them. These 
derived column densities are all $\ga$10$^{20}$~cm$^{-2}$, more than an order 
of magnitude above our observational limit. Assuming that our method of 
inferring H{\sc i} size from optical size is robust, then this result implies 
that there is no significant population of galaxies with H{\sc i} column 
densities $\la$10$^{20}$~cm$^{-2}$. We are currently undertaking a VLA and 
ATCA survey  of all the HIDEEP sources in order to measure their column 
densities directly. This will enable us to test this result without having to 
infer H{\sc i} size from optical size. 

Possible physical explanations for this intriguing second result are discussed
in detail 
elsewhere (Disney \& Minchin 2003). Ionisation is unlikely to be responsible  
since the intergalactic radiation field locally is at least an order to 
magnitude too small (Scott et al. 2002).  Another possibility, raised in the 
early days of 21-cm astronomy, is `freezing out' -- i.e. that the 21-cm 
spin temperature of low column density objects may fall to the cosmic 
background temperature, rendering such clouds invisible in emission. 
 
The cosmic significance of gas-rich  low-surface brightness galaxies will be 
the subject of a subsequent paper based on this same data (Minchin et al. 
2003, in preparation). However, the fact that we have not found a single 
galaxy in H{\sc i} which cannot be seen in our optical data suggests that 
there cannot be large numbers of gas-rich extremely LSB galaxies or 
intergalactic clouds. 

If our second result is confirmed by the follow-up VLA and ATCA observations, 
then this will imply that H{\sc i} surveys need not have column density limits
much lower than $\sim$10$^{20}$~cm$^{-2}$, since few galaxies would appear to
have column densities lower than this. In particular, this would imply that 
previous large-area, shallow surveys, e.g. HIPASS and HIJASS, will not have 
missed many low-column density objects. However, long integrations using 
telescopes with smaller beams would improve the statistics on the numbers of 
low mass, low column-density sources.  

\section*{Acknowledgements}

The authors would like to thank Lister Staveley-Smith, Tony Fairall, David
Barnes, Jon Davies and Suzanne Linder for useful discussions and we are
particularly in debt to Ron Ekers, the ATNF director, for the construction
of the multibeam system.  R. F. Minchin, W. J. G. de Blok and P. J. Boyce 
acknowledge the support of PPARC.  We also thank the anonymous referee whose
comments lead to a drastic revision in the presentation of this paper.
The authors would also like to thank the 
staff of the CSIRO Parkes and Narrabri observatories for their help with 
observations.  We
acknowledge PPARC grant GR/K/28237 to MJD towards the construction of the
multibeam system and PPARC grants PPA/G/S/1998/00543 and PPA/G/S/1998/00620 
to MJD towards its operation.  The Australia Telescope is funded by the
Commonwealth of Australia for operation as a National Facility managed
by CSIRO.

This research has made use of the NASA/IPAC Extragalactic Database (NED) which
is operated by the Jet Propulsion Laboratory, Caltech, under agreement with 
the National Aeronautics and Space Administration.  This research has also 
made use of the Digitised Sky Survey, produced at the Space Telescope Science
Institute under US Government Grant NAG W-2166 and of NASA's Astrophysics 
Data System Bibliographic Services

\appendix
\section{Derivation of the surface-brightness -- column-density relation}
\label{disney-banks-deriv}

 For any given area the H{\sc i} surface density and the
optical surface brightness are related by:
\begin{equation}
\Sigma_{HI} \left(M_\odot \hbox{pc}^{-2}\right) =
\left(\frac{M_{HI}}{L_{B}}\right) \times 
\Sigma_{B} \left(L_\odot \hbox{pc}^{-2}\right)
\end{equation}

\noindent where $\Sigma_{HI}$ is the H{\sc i} surface density and
$\Sigma_{B}$ is the {\it B}-band optical surface brightness, averaged
over the region, and $M_{HI}$ and $L_{B}$ are the H{\sc i} mass
and the {\it B}-band luminosity within the same region.  As 1 $M_\odot$
pc$^{-2}$ is approximately equal to an H{\sc i} column density of
10$^{20.1}$~cm$^{-2}$ and 1 $L_\odot$ pc$^{-2}$ is approximately
equal to a surface-brightness of 27.05~{\it B}\,mag\,arcsec$^{-2}$, this
gives the scaling relationship:
\begin{equation}
N_{HI} \simeq 10^{20.1}\left(\frac{M_{HI}}{L_{B}}\right)
10^{\left(0.4\left(27-\mu_{mean}\right)\right)}
\end{equation}

\noindent where  $N_{HI}$ is the H{\sc i} column density in units of cm$^{-2}$ 
and $\mu_{mean}$ is the average optical surface brightness in units of
mag\,arcsec$^{-2}$ taken over the same area as $N_{HI}$.  This 
can be re-written as:
\begin{equation}
\mu_{mean} \simeq 2.5\left(30.1 + 
      \log\left(\frac{M_{HI}}{L_{B}}\right) - 
      \log\left(N_{HI}\right)\right)
\end{equation}

This relation can be adapted to relate the central surface-brightness
of a galaxy to its average H{\sc i} column density if certain
assumptions are made about the size of the H{\sc i} disc.  Cayatte et
al. (1994) found that $R_{HI} \simeq 1.7 R_{25}$.  This scaling
will obviously not hold for LSB galaxies, some of which may not even
have a $\mu_B = 25$ isophote, but for a Freeman's law galaxy $R_{25} =
3.1$ scale lengths and it therefore seems reasonable to assume that
$R_{HI} = 3.1\times 1.7 = 5.25$ scale lengths.  Assuming also that 
$M_{HI}/L_B = 0.3 M_\odot/L_\odot$ (average value from Roberts \& 
Haynes 1994), we get:
\begin{equation}
\mu_0 \simeq 2.5\left(28.95 - \log\left(N_{HI}\right)\right)
\end{equation}

\noindent which can be used to work out an approximate equivalent
central surface-brightness limit for H{\sc i} surveys.

\section{Column-density sensitivity}
\label{col-density-sens}

The signal entering the receiver is measured in terms of the
antenna temperature $T_A$ where

\begin{equation}
k T_A = S_\nu D^2
\end{equation}

\noindent $S_\nu$ is the strength of the source in flux units and $D$ the
dish-diameter.  For a significant detection $T_A$ should exceed the 
uncertainty in the system power by some
signal-to-noise-ratio $\sigma$, i.e.

\begin{equation}
T_A \geq \sigma \frac{T_{sys}}{\sqrt{t_{int}\Delta \nu}}
\label{anteqn}
\end{equation}

\noindent which is the usual `antenna equation' derived from the bandwidth
theorem.

In the Rayleigh-Jeans regime, surface brightness is conventionally expressed
in terms of the brightness temperature

\begin{equation}
T_B = \frac{\lambda^2}{2 k} \frac{S_\nu}{\Delta \Omega}
\end{equation}

\noindent where $\Delta\Omega$ is the solid angle of the source.  For H{\sc i}
galaxies quantum mechanics yields 

\begin{equation}
\frac{N_{HI}}{\Delta V} = 1.8\times 10^{18}T_B
\end{equation}

\noindent where $N_{HI}$ is the column-density in cm$^{-2}$ and $\Delta V$
is the velocity width of the line.

The antenna equation (\ref{anteqn}) can now be rewritten as

\begin{equation}
\frac{N_{HI}}{\Delta V} > 1.8\times 10^{18} \frac{\lambda^2}{2D^2\Delta\Omega}
\frac{\sigma T_{sys}}{\sqrt{t_{obs}\Delta\nu_{ch}}}
\end{equation}

\noindent where we have replaced $\Delta\nu$ with $\Delta\nu_{ch}$ because
we are here interested in a peak-flux limited survey.  

If we now substitute in the beam size, $\Omega_b = 1.13(\lambda/D)^2$,
and the channel velocity-width $\Delta V_{ch} = (c/\nu_{rest}) \times 
\Delta\nu_{ch}$, we get

\begin{equation}
\frac{N_{HI}}{\Delta V} \geq 5.5\times 10^{16} \frac{\Omega_b}{\Delta\Omega}
\frac{\sigma T_{sys}}{\sqrt{t_{obs} \Delta V_{ch}}}
\end{equation}

So, if the source fills the beam, i.e. $\Omega_b/\Delta\Omega = 1$

\begin{equation}
\frac{N_{HI}}{\Delta V} \geq 5.5\times 10^{16} 
\frac{\sigma T_{sys}}{\sqrt{t_{obs} \Delta V_{ch}}}
\label{diseqnB6}
\end{equation}

\noindent which agrees with Equation \ref{diseqn2}, and 
is indeed independent of
telescope diameter.  For HIDEEP, $T_{sys} = 26$K, $t_{obs} = 9000$s, $\Delta
V_{ch} = 13.2$ km\,s$^{-1}$ and if $\sigma = 5$

\begin{equation}
\frac{N_{HI}}{\Delta V} \geq 2.1\times 10^{16} 
\hbox{cm$^{-2}$/km\,s$^{-1}$}
\end{equation}

For an optimally smoothed survey limited only by the signal-to-noise of the
total flux of the galaxy rather than the flux in a single channel,
one would replace $\Delta 
V_{ch}$ by $(\Delta V/c)\nu_{21}$ in Eqn \ref{diseqnB6} and reach (Disney
\& Banks 1997)

\begin{equation}
\left(\frac{N_{HI}}{\Delta V}\right) > 1.8\times 10^{18} \sigma T_{sys}
\sqrt{\frac{1}{\Delta V t_{obs}}}
\end{equation}

\section{Calculating minimum column-density sensitivity from the lowest 
flux sources actually detected}
\label{col-density-calc}

At any distance a source that is just detectable because of its H{\sc i}
mass, and which just fills the beam, will have the minimum column-density
the survey is capable of detecting:

\begin{equation}
N_{HI} = 5\times 10^{20} \frac{M_{HI}}{\pi \theta^2 d_{pc}^2}
\end{equation}

\noindent where $\theta$ is angular diameter of the source in radians, and
$d_{pc}$ its distance in parsecs.  By a well known relation $M_{HI}$ is 
related to flux $F_{HI}$
by:

\begin{equation}
M_{HI} = 2.356 \times 10^5 F_{HI} d_{Mpc}^2
\end{equation}
\noindent and therefore we can substitute this into the equation above to
give

\begin{equation}
N_{HI} = \frac{2.356 \times 10^5 F_{HI} d_{Mpc}^2}{\pi \theta^2 d_{pc}^2}
\end{equation}

\noindent therefore

\begin{equation}
N_{HI}^{min} = 4.5\times 10^{20} \frac{F_{HI}^{min}}{\theta^{\prime 2}}
\end{equation}

\noindent where $\theta^\prime$ is the source diameter in arcminutes.
Putting $\theta^\prime$ equal to the beam diameter, and $F_{HI}$ to the
lowest flux measured in the survey will yield the column-density limit
of the survey at the velocity width of that source.  Of more interest is 
putting in the lowest value of $F_{HI}/\Delta V$, which allows the 
column-density limit to be found at all velocity widths:

\begin{equation}
\left(\frac{N_{HI}}{\Delta V}\right)^{min} = 4.5\times 10^{20} 
\left(\frac{F_{HI}}{\Delta V}\right)^{min}\frac{1}{\theta^{\prime 2}}
\end{equation}

\noindent which is Eqn \ref{nhieqn2} in the text.
\end{document}